\documentclass[a4paper,11pt]{article}
\pdfoutput=1 

\usepackage{jheppub} 
\usepackage{slashed}
\usepackage{subfigure}
\usepackage{rotating}
\usepackage{longtable}

\title{The case for 100 GeV bino dark matter:\\ A dedicated LHC tri-lepton search}

\author[a]{Melissa van Beekveld,}
\author[a,b]{Wim Beenakker,}
\author[a,c]{Sascha Caron}
\author[d]{Roberto Ruiz de Austri}

\affiliation[a]{Institute for Mathematics, Astrophysics and Particle Physics, Radboud University Nijmegen, Heyendaalseweg 135, Nijmegen, The Netherlands}
\affiliation[c]{Nikhef, Science Park 105, Amsterdam, The Netherlands}
\affiliation[b]{Institute  of  Physics,  University  of  Amsterdam,  Science  Park  904,  Amsterdam,  The
Netherlands}
\affiliation[d]{Instituto de Fisica Corpuscular, IFIC-UV/CSIC, Calle Catedrático José Beltran 2, Valencia, Spain}

\emailAdd{mcbeekveld@gmail.com}
\emailAdd{w.beenakker@science.ru.nl}
\emailAdd{scaron@cern.ch}
\emailAdd{rruiz@ific.uv.es}

\abstract{Global fit studies performed in the pMSSM and the photon excess signal originating from the Galactic Center seem to suggest compressed electroweak supersymmetric spectra with a $\sim$100 GeV bino-like dark matter particle. We find that these scenarios are not probed by traditional electroweak supersymmetry searches at the LHC. We propose to extend the ATLAS and CMS electroweak supersymmetry searches with an improved strategy for bino-like dark matter, focusing on chargino plus next-to-lightest neutralino production, with a subsequent decay into a tri-lepton final state. We explore the sensitivity for pMSSM scenarios with $\Delta m = m_{\rm NLSP} - m_{\rm LSP} \sim (5 - 50)$ GeV in the $\sqrt{s} = 14$ TeV run of the LHC. Counterintuitively, we find that the requirement of low missing transverse energy increases the sensitivity compared to the current ATLAS and CMS searches. With 300 fb$^{-1}$ of data we expect the LHC experiments to be able to discover these supersymmetric spectra with mass gaps down to $\Delta m \sim 9$ GeV for DM masses between 40 and 140 GeV. We stress the importance of a dedicated search strategy that targets precisely these favored pMSSM spectra.} 


\begin{document} 
\maketitle
\flushbottom

\section{Introduction}

The existence of dark matter (DM) is widely accepted, but its fundamental nature remains unknown. The leading theory is that DM consists of weakly interacting massive particles (WIMPs), i.e.~particles that have no electromagnetic or color charge. WIMPs are particularly favored due to the \emph{WIMP miracle}: weak-scale particles (with masses around 100-1000~GeV) can result in a DM relic density that is consistent with the value provided by the Planck collaboration ($\Omega h^2 = 0.118$~\cite{Planck:2015}). WIMPs can be detected directly and indirectly. Direct detection methods aim to measure nuclear recoils that originate from collisions between WIMPs and the target material of the detector (for a review, see for instance ref.~\cite{Undagoitia:2015gya}). Indirect detection methods try to observe annihilation products of WIMPs (for a review, see for instance ref.~\cite{BS:2010}). These methods focus on locations of high DM density, such as the center of the Milky Way or dwarf spheroidal satellite galaxies (dSphs) of the Milky Way. Observations of the center of our Galaxy with the Large Area Telescope (LAT), aboard the Fermi satellite, show a photon excess emanating from this region~\cite{Goodenough:2009gk, Vitale:2009hr, Hooper:2010mq, Hooper:2011ti, Abazajian:2012pn,Gordon:2013vta, Hooper:2013rwa, Abazajian:2014fta, Daylan:2014rsa,Calore:2014xka,TheFermi-LAT:2015kwa}. \\
A theoretical framework for WIMPs can be provided by supersymmetry (SUSY). This theory postulates for each Standard Model (SM) particle the existence of a superpartner (or \emph{sparticle}) state whose spin differs by $1/2$. In the $R$-parity conserving phenomenological version of the minimal supersymmetric standard model (pMSSM),the introduction of these new sparticles can provide a solution to the hierarchy problem as well as WIMPs, for example the lightest neutralino ($\tilde{\chi}^0_1$), which is a DM candidate when it is the lightest supersymmetric particle (LSP). It has recently been shown that the annihilation of $\tilde{\chi}^0_1$ pairs in the pMSSM framework is a possible explanation for the Galactic Center (GC) photon excess \cite{Caron:2015wda, 2015arXiv150707008B}. The best fit to the data corresponds to pMSSM models with mostly bino-like LSPs with masses $m_{\tilde{\chi}^0_1} \sim$80-90 GeV and mostly higgsino- or wino-like next-to-lightest neutralino and chargino states (NLSPs) with a mass close to the lightest neutralino mass. Furthermore, these same models are also consistent with a small photon excess observed in the dwarf galaxy Reticulum~II~\cite{Geringer-Sameth:2015lua, 2015JCAP...12..013A} and Tucana~III~\cite{Li:2015kag}.   \\
In addition, global fit studies performed in the pMSSM with 15 parameters suggest a bino-like LSP with $m_{\tilde{\chi}^0_1} \sim $100~GeV~\cite{Strege:2014ija, 2015arXiv150707008B}. These studies are performed including all available accelerator, direct-detection and cosmology constraints and the GC photon excess. An analysis of the parameter space of the pMSSM with 10 parameters (pMSSM10), including constraints from the Higgs mass, B-meson observables, electroweak precision observables, the DM relic density and spin-independent DM scattering, shows that the most likely pMSSM10 models have a bino-like LSP with a mass around 100-200 GeV. Furthermore, the mass difference between the LSP and heavier neutralino and chargino NLSP states in these models is $~20$ GeV at most at low LSP masses~\cite{2015EPJC...75..422D, Bagnaschi:2015eha}. \\
\begin{figure}[b]

	\begin{center}
\includegraphics[width=0.7\textwidth]{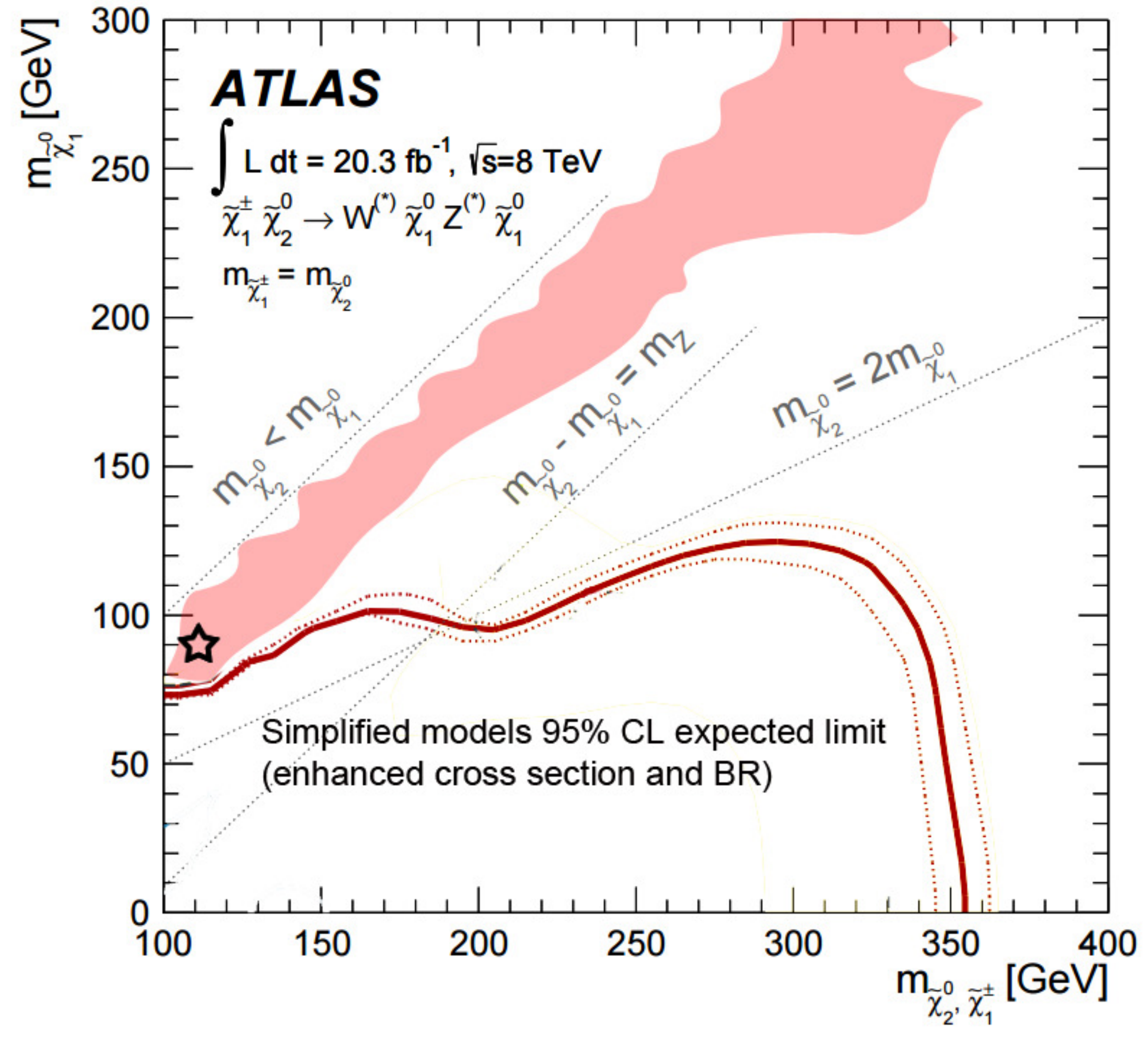}
\caption{The 95$\%$ confidence level exclusion limit on direct production of $\tilde{\chi}^{\pm}_1\tilde{\chi}^0_2$  with $WZ$-mediated decays~\cite{ATLASLIM}. This limit is obtained using simplified models, where the NLSPs are assumed to be 100$\%$ wino-like and relevant branching ratios are set at $100\%$. The star indicates the GC best fit pMSSM models from ref.~\cite{Caron:2015wda}, which coincide with the best global fit models obtained by~\cite{2015arXiv150707008B}. These models will not have NLSPs that are 100$\%$ wino-like, which reduces the production cross section and the relevant branching ratios.  The shaded red area indicates the $1\sigma$ contour of the most likely pMSSM10 models from ref.~\cite{Bagnaschi:2015eha}. } 
\label{fig:EWlim}
	\end{center}
\vspace{-2cm}
\end{figure} 

To conclude, some signs coming from independent analyses justifies further studies on pMSSM scenarios with a $\sim$100 GeV bino-like DM particle and a $\sim$10-25 GeV heavier chargino and neutralino. This motivates a dedicated search at the Large Hadron Collider (LHC) for such weakly-interacting particles with masses that could reside at or near the weak scale. Electroweak SUSY searches at the LHC are typically performed using multi-lepton search channels, where the leptons originate from the decay of pair produced charginos and neutralinos. Typical search channels look for signatures that include same- or opposite-sign di-leptons, tri-leptons, four leptons and a large missing transverse energy ($\slashed{E}_T$) \cite{ATLAS:2012uks, Aad:2012hba, Aad:2012pxa, ATLAS:2013rla, 2014PhRvD..89i2007C, 2014JHEP...05..071A,2014EPJC...74.3036K, 2014arXiv1407.0350A, Aad:2014nua, 2015EPJC...75..208A, 2015arXiv150907152A, 2015arXiv150702898C}. Previous searches for electroweak SUSY production at the LHC found no significant excess. The LHC experiments have been able to constrain electroweak sparticle masses, but the existing search techniques fail when the mass differences between the LSP and the NLSPs become too small (figure~\ref{fig:EWlim}). Standard searches for multi-lepton plus $\slashed{E}_T$ signals rely on triggers that require $p_{T}(l) > 20$ GeV for the transverse momentum of a lepton. The energy of the produced leptons is roughly bounded by $(m_{\rm NLSP}- m_{\rm LSP})/2$, therefore searches start to lose sensitivity when the mass differences drop below $~40-50$ GeV. We find that, even with the high-luminosity upgrade of the LHC resulting in an integrated luminosity of 3000 fb$^{-1}$, the LHC experiments will not be sensitive to these very important dark matter scenarios using their current tri-lepton search strategies (see figure~\ref{fig:winoatl3000}, page~\pageref{fig:winoatl3000}). We therefore stress the importance of a dedicated search strategy that targets precisely these pMSSM spectra. \\

\begin{figure}[h]
	\begin{center}
		\includegraphics[width=0.6\textwidth]{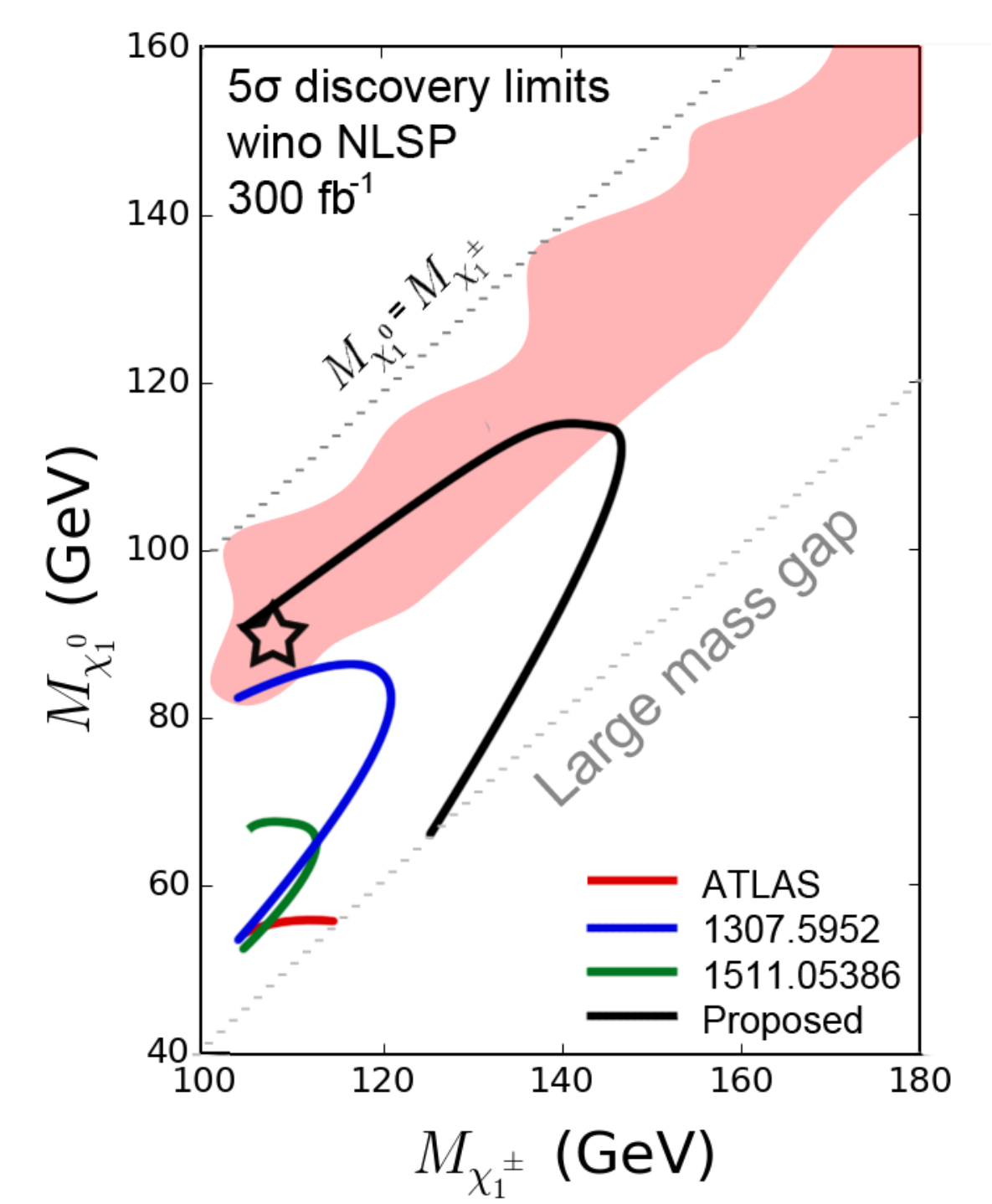}
		\caption{The expected $5\sigma$ discovery reach for the LHC at 14 TeV and with 300 fb$^{-1}$ of data for pMSSM models with a bino-like LSP and wino-like NLSPs, assuming a background uncertainty of $10\%$. The black line indicates the reach for the proposed analysis. The blue line and green line indicate the reach obtained for ref. \cite{Gori:2013ala} and ref.~\cite{Cao:2015efs} respectively. The red line indicates the current reach for the ATLAS tri-lepton analysis (ref.~\cite{2014arXiv1402.7029A}). The cuts used to create this figure are shown in table~\ref{tab:searches}. Ref.~\cite{Schwaller:2013baa} is not included, because their analysis did not reach $5\sigma$ for these pMSSM models. The star indicates the GC best fit pMSSM models from ref.~\cite{Caron:2015wda}, which coincide with the best global fit models obtained by~\cite{2015arXiv150707008B}.  The shaded red area indicates the $1\sigma$ contour of the most likely pMSSM10 models from ref.~\cite{Bagnaschi:2015eha}. } 
		\label{fig:existing}
	\end{center}
\end{figure}

This paper addresses these pMSSM scenarios in wino-/higgsino-like chargino and neutralino production at the LHC with $\sqrt{s} = 14$ TeV, which result in a tri-lepton plus missing transverse energy ($3l + \slashed{E}_T$) final state. We investigate the role of the missing transverse energy and lepton transverse momentum in the search for electroweak SUSY with mass splittings $\Delta m \equiv m_{\rm NLSP}- m_{\rm LSP} \sim (5 - 50)$ GeV. We will offer an improved search strategy for the $3l + \slashed{E}_T$ channel, which extends the exclusion reach for the compressed pMSSM models tremendously. \\
A lot of work has already been done to gain sensitivity in similar pMSSM scenarios. The use of a hard initial state, for example a jet (e.g. ref.~\cite{Dreiner:2012gx, Berggren:2013vfa, Delgado:2012eu}) or a photon (e.g. ref.~\cite{Abbiendi:2002vz, Heister:2002mn, Abreu:2000as}) has been suggested. The use of soft leptons in combination with a jet has been suggested as well (e.g. ref.~\cite{Zarzhitsky:1972078, Gori:2013ala, Schwaller:2013baa}). We investigated the sensitivity of these searches for our models. Since the considered models have a large bino component ($\sim$90$\%$), the standard mono-jet and mono-photon searches (where two LSPs are produced) will not be sensitive due to the small production cross section of LSPs. Furthermore, the sfermion masses  are all set at the multi-TeV scale in this analysis, so the t-channel squark exchange channel is suppressed due to high squark masses. We also investigated the possibility of LSP production via vector boson fusion, but only 3 events are expected at 300 fb$^{-1}$. We therefore decided to focus on the tri-lepton search channel. We found that by demanding an extra photon or jet, the production cross section is reduced by a factor 10. In table~\ref{tab:searches} a short summary is given of the cuts that are used in some of the existing or proposed tri-lepton searches. We find that these searches are not sensitive to the pMSSM models favored by the GC excess photon spectrum and global fits (figure~\ref{fig:existing}). \\ 

\clearpage
The paper is organized as follows. In section~\ref{sec:theo} we will provide a brief overview of the theoretical background. In section~\ref{sec:signal} we will present the details of the Monte Carlo simulation of the signal and background processes considered. In section~\ref{sec:dis} we will look at discriminating parameters. Finally, in section~\ref{sec:res} we will present our results.
\clearpage
\begin{longtable}{| l | l | }
\hline
search & signal selection~\\
\hline
\hline
& ATLAS and CMS tri-lepton searches at 8 TeV\\
\hline
\hline
ATLAS~\cite{2014arXiv1402.7029A}  & single $e$ and $\mu$ trigger: $p_T(l_1) > 25$ GeV\\
& symmetric di-muon trigger: $p_T(\mu_1)$ and $p_T(\mu_2) > 14$ GeV\\
& asymmetric di-muon trigger: $p_T(\mu_1) > 18$ GeV and $p_T(\mu_2) > 10$ GeV \\
& symmetric di-electron trigger: $p_T(e_1)$ and $p_T(e_2) > 14$ GeV \\
& asymmetric di-electron trigger: \\
& \hspace{2cm}$p_T(e_1) > 25$ GeV and $p_T(e_2) > 10$ GeV\\
& electron-muon (muon-electron) combi trigger:\\
& \hspace{2cm}$p_T(e_1) > 14 (10)$ GeV and $p_T(\mu_1) > 10 (18)$ GeV\\
& at least one OSSF\footnote{Opposite Sign Same Flavor} lepton pair with $12 < M_{l^+ l^-} < 60 $ GeV \\
& $\slashed{E}_T > 50$ GeV \\
& $p_T(l_3) > 10$ GeV \\
\hline
CMS~\cite{Khachatryan:2014qwa}  & single $e$ and $\mu$ trigger: $p_T(e) > 27$ GeV or $p_T(\mu) > 24$ GeV\\
& di-muon or di-electron or combination: $p_T(l_1) > 20$ and $p_T(l_2) > 10$ GeV \\
& at least one OSSF lepton pair with $12 < M_{l^+ l^-} < 75 $ GeV \\
&$\slashed{E}_T > 50 $ GeV\\
&$p_T(l_3) > 8$ GeV\\
\hline
\hline
& soft tri-lepton searches (theory prospects) \\
\hline
\hline
1312.7350 \cite{Schwaller:2013baa} & only allow for soft leptons $5 < p_T(\mu) < 20$ GeV \\
& \hspace{2cm} and $10 < p_T(e) < 20$ GeV (veto on higher $p_T$ leptons) \\
& $\slashed{E}_T > 300-1000 $ GeV and $p_T(j_1) > 300-1000 $ GeV (50 GeV steps)\\
& $\Delta\phi(j_1, j_2) < 2.5$ \\
& $p_T(j_3) < 30$ GeV  \\
\hline
1307.5952 \cite{Gori:2013ala} & exactly 3 leptons with $ 7 < p_T(l) < 50$ GeV\\
& at least one OSSF lepton pair with $12 < M_{l^+ l^-} < 30-50$ GeV \\
& initial state radiation jet with $p_T(j) > 30 $ GeV and within $|\eta(j)| < 2.5$ \\
& $\slashed{E}_T > 60 $ GeV\\
& $p_T(l_1)/p_T(j_1) < 0.2$ and $\slashed{E}_T/p_T(j_1) < 0.9$\\
\hline
1511.05386 \cite{Cao:2015efs} &  exactly 3 leptons recorded with any of the ATLAS lepton triggers\\
& at least one OSSF lepton pair with $12 < M_{l^+ l^-} < 40$ GeV \\
& $\slashed{E}_T > 50 $ GeV\\
\hline

\caption{Summary of cuts used/proposed in various tri-lepton searches at the LHC.} 

\label{tab:searches}
\end{longtable}

\newpage

\section{Theoretical background}
\label{sec:theo}

Charginos and neutralinos are the mass eigenstates of the superpartners of the electroweak   gauge bosons (\emph{bino} and \emph{wino}) and the two Higgs doublets (\emph{higgsinos}). These particles mix under the influence of electroweak symmetry breaking. The neutral states mix as a result of the non-diagonal neutralino mass matrix:\\

\begin{center}
$\textbf{M}_{\tilde{\chi}^0} = \begin{pmatrix}
 M_1 & 0 & -c_{\beta}s_{\theta_W}M_Z & s_{\beta}s_{\theta_W}M_Z \\
 0 & M_2 & c_{\beta}c_{\theta_W}M_Z & -s_{\beta}c_{\theta_W}M_Z \\
 -c_{\beta}s_{\theta_W}M_Z & c_{\beta}c_{\theta_W}M_Z  & 0 & -\mu  \\
 s_{\beta}s_{\theta_W}M_Z & -s_{\beta}c_{\theta_W}M_Z & -\mu & 0 
\end{pmatrix}$
\end{center}

\noindent where $M_1$, $M_2$ and  $\mu$ are the bino, wino and Higgsino masses.  The ratio of the vacuum expectation values of the two Higgs doublets is denoted by $\tan\beta$ and $M_Z$ is the $Z$ boson mass. The cosine and sine of the weak mixing angle $\theta_W$ are indicated by $c_{\theta_W}$ and $s_{\theta_W}$. Following the same notation, $c_{\beta}$ and $s_{\beta}$ indicate the cosine and sine of $\beta$. The chargino mass matrix is given by:
\begin{center}
$\textbf{M}_{\tilde{\chi}^{\pm}} = \begin{pmatrix}
 M_2 & \sqrt{2}c_{\beta}M_W \\
 \sqrt{2}s_{\beta}M_W & \mu 
 \end{pmatrix}$
\end{center}
where $M_W$ is the $W$ boson mass. After diagonalization of these mass matrices, the mass eigenstates will be labeled as $\tilde{\chi}^{0}_{1,2,3,4}$ and $\tilde{\chi}^{\pm}_{1,2}$, in increasing mass order. We demand that $\tilde{\chi}^0_1$ is the LSP. This particle is stable, provided that $R$-parity is conserved, and weakly interacting, which makes it a WIMP candidate. \\

The amount of bino, wino and higgsino mixing is controlled by the mass hierarchy of the interaction eigenstates ($M_1$, $M_2$ and $\mu$). If the mass difference between the interaction eigenstates is big ($\gg M_W$), mixing will be suppressed. In that case, simplified models~\cite{2009PhRvD..79g5020A} can be obtained, which have successfully been invoked by the LHC experiments to set constraints on neutralino and chargino masses. In figure~\ref{fig:EWlim} we show the ATLAS limits obtained in these simplified models. The branching ratios for the indicated decay channels are set at $100\%$ and the NLSPs are assumed to be $100\%$ wino.\\ 

\begin{figure}[b]
	\begin{center}
		\includegraphics[width=0.5\textwidth]{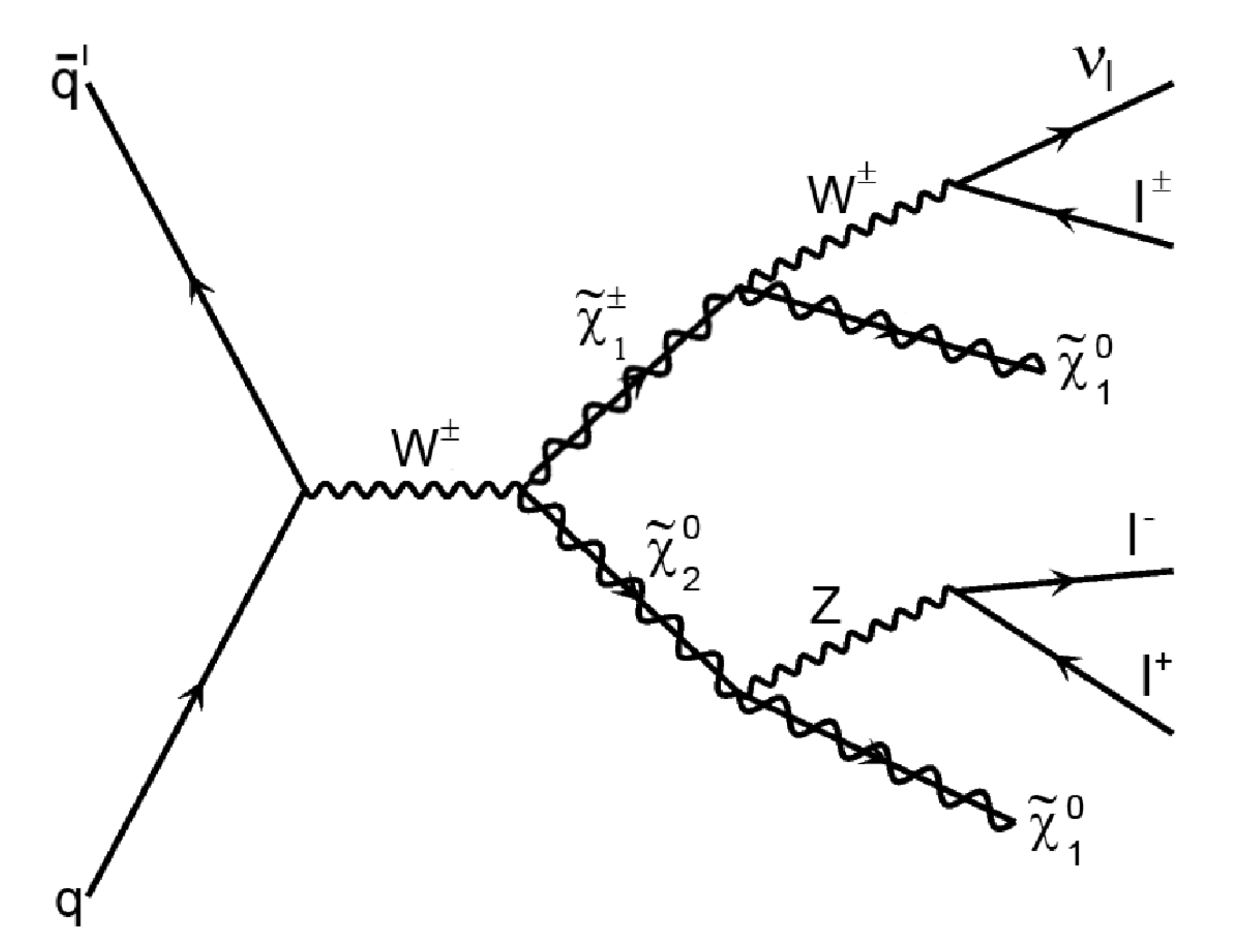}
		\caption{Chargino-neutralino production and decay to tri-lepton final states via gauge bosons in the LHC.} 
		\label{fig:feynman}
	\end{center}
\end{figure}  
In this analysis we will assume pMSSM models with a bino-like LSP (60-99 $\%$), meaning that $M_1 < M_2, |\mu|$. We will consider two NLSP configurations. In one configuration, we assume wino-like NLSPs (80-99 $\%$) with $m_{\tilde{\chi}^0_2} \sim m_{\tilde{\chi}^{\pm}_1}$, $\Delta m \sim (5-50)$ GeV and $m_{\tilde{\chi}^{\pm}_2}, m_{\tilde{\chi}^{0}_{3,4}}$ much heavier (corresponding to $M_1 < M_2 \ll |\mu|$ ). For the other configuration, higgsino-like NLSPs (70-90 $\%$) are assumed, with $m_{\tilde{\chi}^0_{2,3}} \sim  m_{\tilde{\chi}^{\pm}_1}$, $\Delta m \sim (5-50)$ GeV and $m_{\tilde{\chi}^{\pm}_2}, m_{\tilde{\chi}^{0}_{4}}$ much heavier (corresponding to $M_1 < |\mu| \ll M_2$). We will assume that all squark and slepton masses are at the multi-TeV scale. We will refer to these two configurations as \emph{wino} NLSP and \emph{higgsino} NLSP respectively.  \\

Slepton-mediated production processes and decays will be suppressed in these pMSSM scenarios. The charginos and neutralinos therefore predominantly decay via off-shell gauge bosons as:
\begin{equation}
\tilde{\chi}^{\pm}_1 \tilde{\chi}^0_2 \rightarrow l^{\pm} \nu \tilde{\chi}^0_1 l^+l^-\tilde{\chi}^0_1, 
\end{equation}
where $l = \mu, e$. The dominant production and decay channel is illustrated in figure~\ref{fig:feynman}.  The corresponding branching ratios are given by: ${\rm BR}(\tilde{\chi}^0_2 \rightarrow l^+l^-\tilde{\chi}^0_1) = 0.07$ and ${\rm BR}(\tilde{\chi}^{\pm}_1 \rightarrow l^{\pm}\nu\tilde{\chi}^0_1) = 0.22$, and the total branching ratio by ${\rm BR}(\tilde{\chi}^{\pm}_1 \tilde{\chi}^0_2 \rightarrow l^{\pm}  l^+l^-\slashed{E}_T) \simeq 1.5 \% $.   \\
The LEP experiments searched for SUSY using the $e^{+}e^{-} \rightarrow \tilde{\chi}\tilde{\chi}'$ production processes and provided limits on the invisible $Z$ boson decay width: $\Gamma_{\rm inv} < 3.2$ MeV. The latter imposes a limit of $m_{\tilde{\chi}^0_1} \gtrsim 45$ GeV, unless the LSP has a very small coupling to the $Z$ boson (in that case, the LSP does not have a sizable higgsino component). Searches for charginos and heavier neutralinos resulted in $m_{\tilde{\chi}} \gtrsim 91.9 - 103.5$ GeV, depending on the mass difference between these particles and the LSP~\cite{Benelli:2003mq, Carena:2003aj}. We will use these limits in our SUSY models. We do not demand that our models satisfy limits originating from DM detection experiments or astrophysical experiments (concerning the DM relic density, DM annihilation cross section, or spin-dependent and spin-independent DM-nucleus cross section), in order to consider all regions of parameter space that might be interesting for compressed SUSY scenarios from a particle collider point of view. \\

We will not address pMSSM scenarios where the LSP is wino-like and the NLSPs are higgsino-like as these spectra do not not have a $\tilde{\chi}^0_2$ with a mass close to $\tilde{\chi}^0_1$ when $m_{\tilde{\chi}^0_1}\sim$100~GeV. Scenarios where the LSP and the lightest chargino are wino-like and the next-to-lightest neutralino is bino-like have a very small ($\sim  \mathcal{O}(10^{-5})$) branching ratio of $\tilde{\chi}^0_2 \rightarrow  \tilde{\chi}^0_1 l^+l^-$ and will therefore not be interesting for the considered $3l+\slashed{E}_T$ search channel. Scenarios where the LSP and the NLSPs are higgsino-like typically have a production cross section that is much smaller than in the case of the NLSPs being wino-like.


\section{Background processes}
\label{sec:signal}

The dominant irreducible SM background to the $3l+\slashed{E}_T$ final state is the production of $W+Z$ bosons that decay leptonically. All (irreducible and reducible) background processes that we consider are described below. In all cases only leptonic decays of the gauge bosons are considered. 
\begin{itemize}
\item \textbf{WZ$\boldsymbol{^{*}/\gamma^{*}}$: } This is the main irreducible background under consideration. This process includes all processes mediated by on or off-shell $Z$ bosons as well as photons. This background will have a resonance in the distribution of $M_{l^+l^-}$ at $M_{l^+l^-}$ close to the $Z$-boson mass, and at $M_{l^+l^-} \lesssim 10$ GeV originating from $J/\psi$ mesons, $\Upsilon$ mesons and low-mass Drell-Yan processes.
\item \textbf{WW: } This process contains two leptons and missing transverse energy due to the escaping neutrinos. A third lepton may be faked by initial state radiation (ISR). 
\item \textbf{ZZ: } This process has two or four final state leptons. Missing transverse energy can originate from neutrinos (in the case of two final state leptons) or it can be provided by decays of $\tau$ leptons to neutrinos and lighter leptons. 
\item \textbf{Zb: } Two leptons arise from $Z$ decay and a third lepton may originate from a semi-leptonic bottom quark decay.  
\item \textbf{Wt: } One lepton and missing energy originate from a leptonic $W$ decay, other leptons may originate from a top quark decay or initial state radiation.
\item \textbf{Z$\boldsymbol{\gamma}$: } Two leptons arise from a leptonic $Z$ decay, a third may be faked by a photon. There would be a minimal amount of missing energy in these events. 
\item \textbf{$\boldsymbol{t\bar{t}}$: } Two leptons come from semi-leptonic decays of the top quarks. An additional lepton can enter from various processes like initial state radiation, $b$ decay or it may be faked by a jet. 
\item \textbf{WWW: } Three leptons and three neutrinos will arise from leptonic $W$-boson decays.
\end{itemize}
Background processes that we do not consider include other tri-boson processes and $t\bar{t}$ production in association with a $Z$ or $W$ boson. These processes will generally have small cross sections (\cite{ATLAS:2013rla}) and must be accompanied by ISR or fake identified leptons to match the signal topology. \\

We model all signal and background processes using MadGraph5~\cite{Alwall:2011uj}, using Pythia 8.1~\cite{Sjostrand:2007gs} for parton showering. We allow up to one additional parton in the hard matrix element and adopt the MLM matching scheme~\cite{2008EPJC...53..473A} to avoid double counting. Jets are clustered using the anti-$k_T$ algorithm as implemented in FastJet 3.1.3~\cite{Cacciari:2011ma}. We use \textsc{Delphes}~3~\cite{deFavereau:2013fsa} as a fast detector simulation. The SUSY processes that are considered typically have small $K$-factors from next-to-leading order corrections ($\sim$1.3~\cite{1996hep.ph...11232B}). Previous studies on tri-lepton channels reported $K$-factors for the background processes to be of order unity~\cite{Schwaller:2013baa}. We adopt a conservative approach by not considering the NLO corrections.

\section{Distinct kinematic features of the signal}
\label{sec:dis}

\begin{figure}[h]
\centering

\centering
 \subfigure[]{\label{fig:mlla}\includegraphics[width=0.48\textwidth]{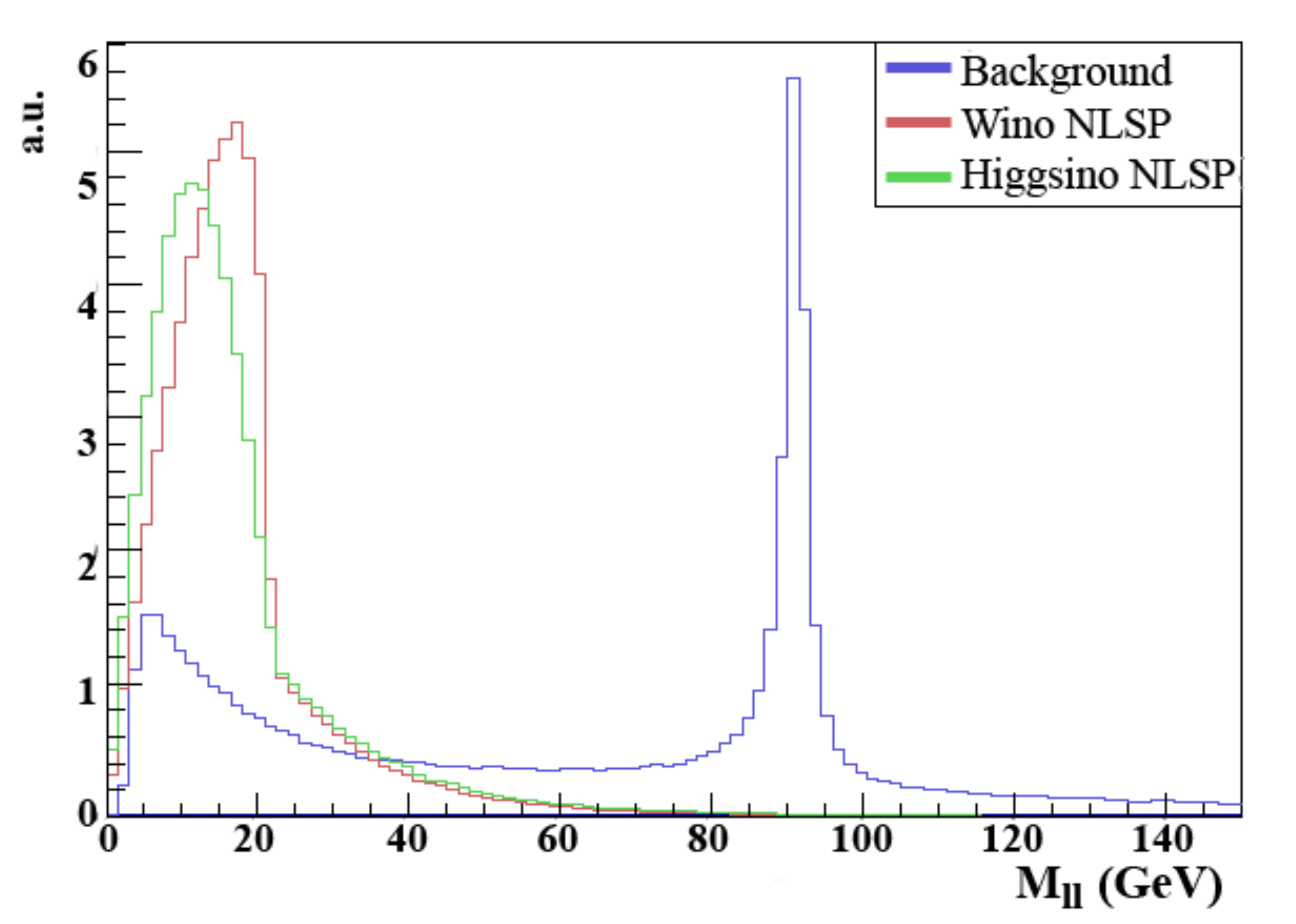}}
 \subfigure[]{\label{fig:metb}\includegraphics[width=0.47\textwidth]{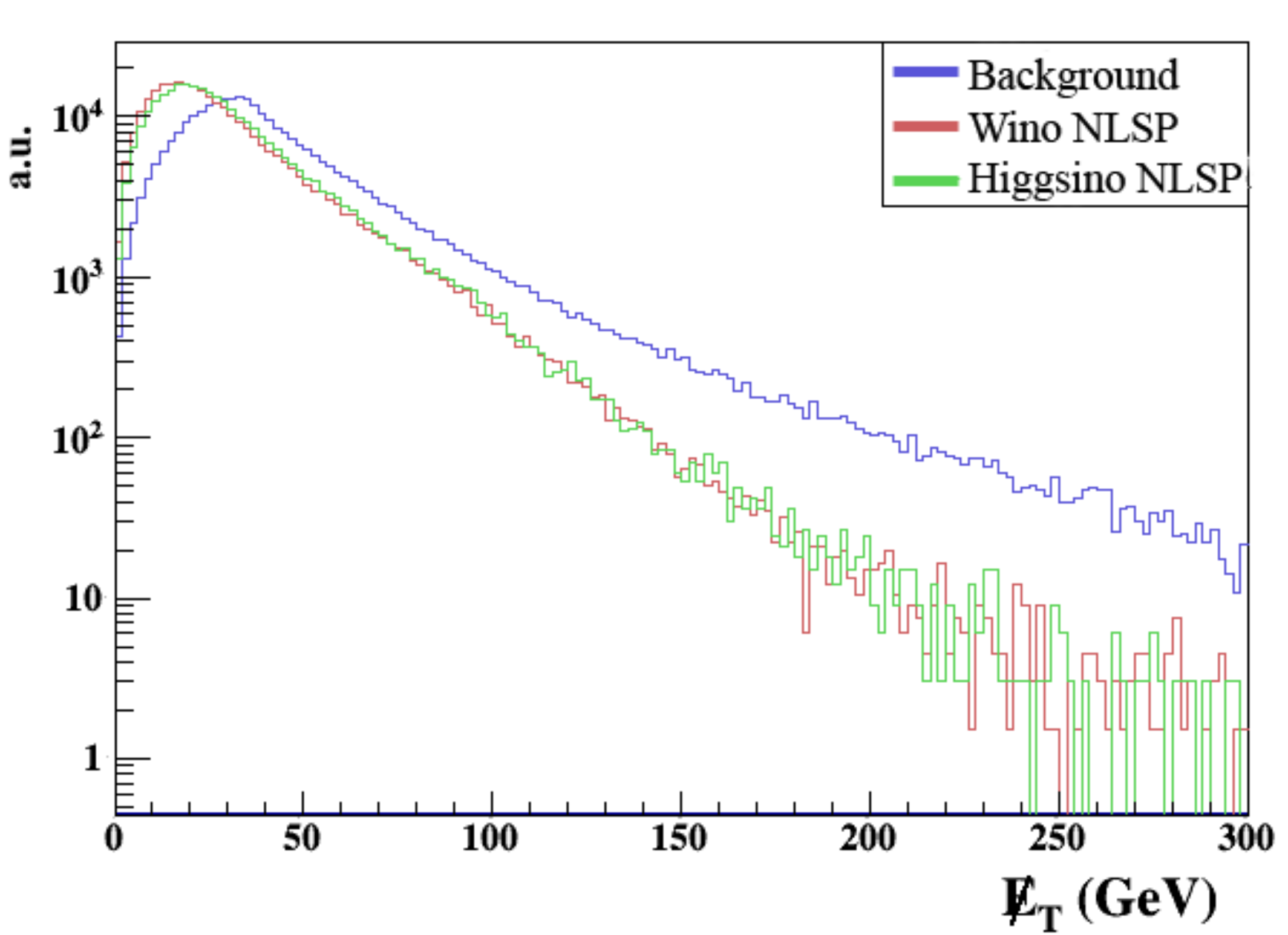}}
\caption{Distribution for $M_{l^+ l^-}$ (\emph{left}) and $\slashed{E}_T$ (\emph{right}) before detector simulation. The blue curve represents the irreducible diboson ($WZ$) background, the red curve represents the wino NLSP scenario and the green curve the higgsino NLSP scenario. In both scenarios, $m_{\tilde{\chi}^0_2} \simeq m_{\tilde{\chi}^{\pm}_1} \simeq 110$ GeV, $m_{\tilde{\chi}^0_1} \simeq 90$ GeV such that $\Delta m \simeq 20$ GeV. All events are normalized to a cross section of 1 pb.}

\centering
 \subfigure[]{\label{fig:diffmassmlla}\includegraphics[width=0.47\textwidth]{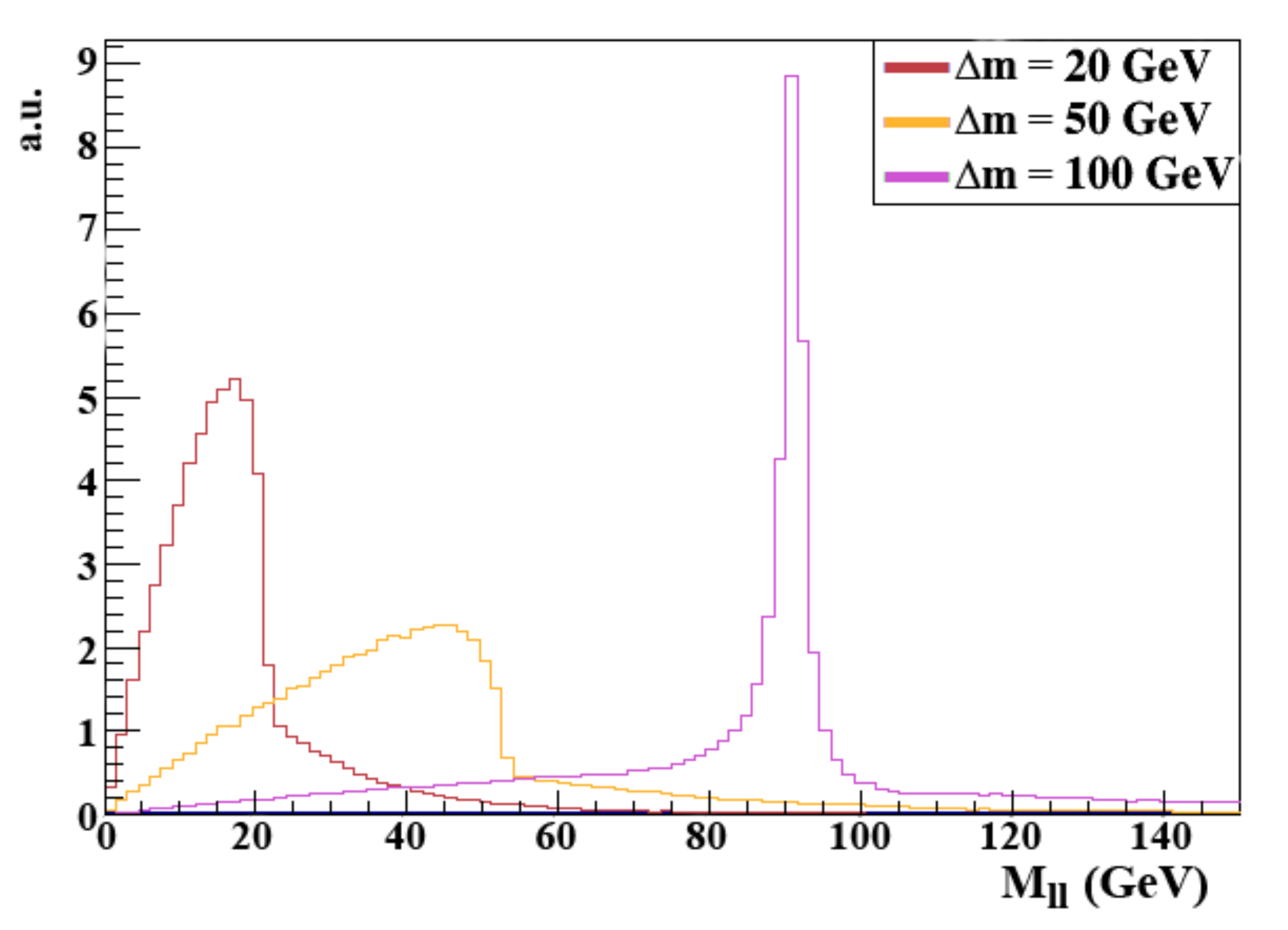}}
 \subfigure[]{\label{fig:diffmassmetb}\includegraphics[width=0.49\textwidth]{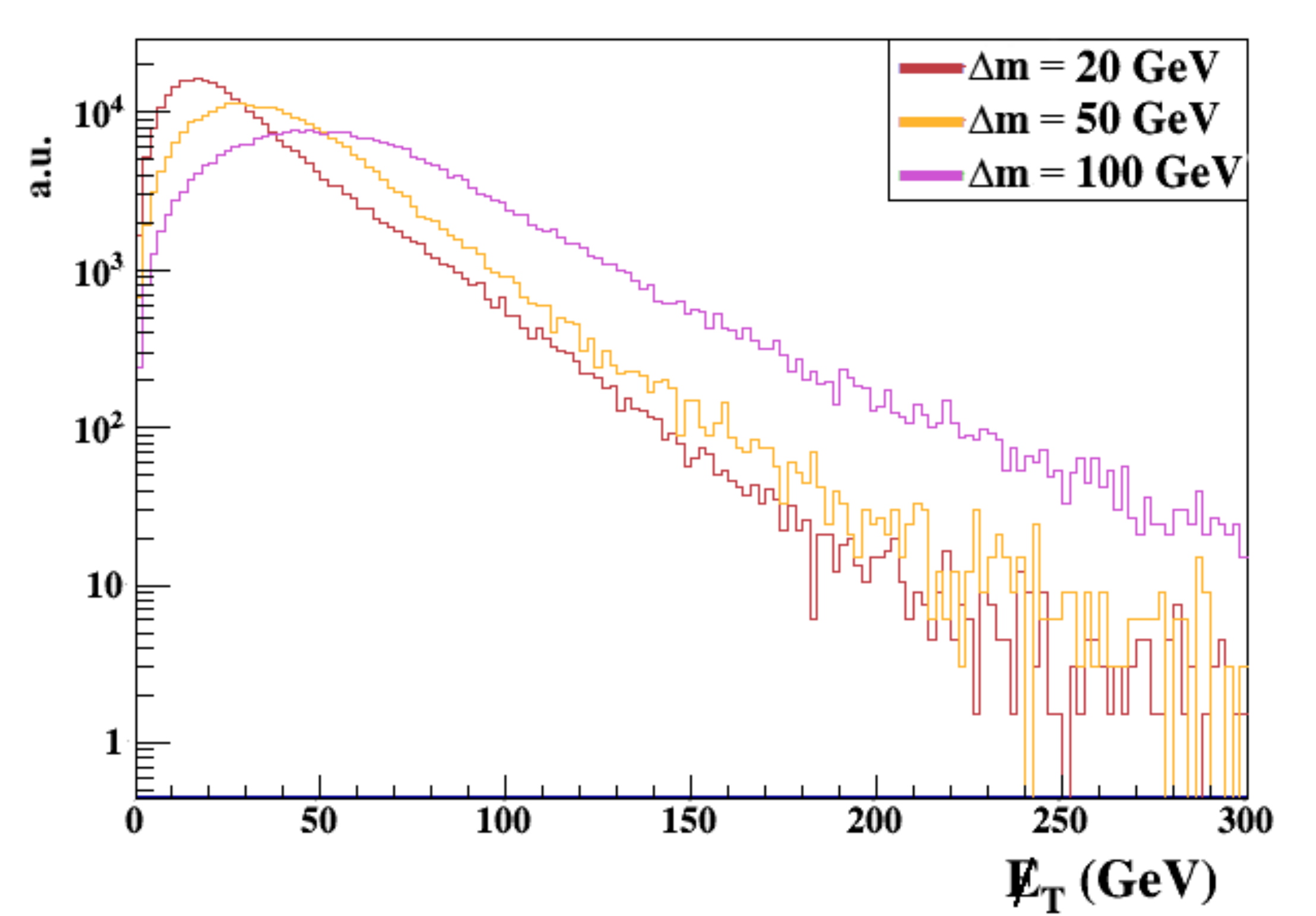}}
\caption{Distribution for $M_{l^+ l^-}$ (\emph{left}) and $\slashed{E}_T$ (\emph{right}) before detector simulation. All three curves represent a wino NLSP, the red curve represents $\Delta m \simeq 20$ GeV, the yellow curve represents $\Delta m \simeq 50$ GeV and the purple curve represents $\Delta m \simeq 100$ GeV. The curves are normalized to a cross section of 1 pb. }
\end{figure}

To evaluate which kinematic features can be invoked to distinguish between signal and background, we will first consider four `benchmark' pMSSM models and the irreducible $WZ$ background. We define these benchmark SUSY models as: three wino NLSP pMSSM models with $\Delta m = 20$ GeV, $50$ GeV and $100$ GeV and one higgsino NLSP pMSSM model with $\Delta m = 20$ GeV. In each model, the LSP has a mass of $m_{\tilde{\chi}^0_1}\sim$90 GeV.\\
The mass gap is clearly visible in the invariant mass of the opposite sign same flavor (OSSF) lepton pair ($M_{l^+ l^-}$) distribution (figure~\ref{fig:mlla} and~\ref{fig:diffmassmlla}), as $M_{l^+l^-}$ is kinematically suppressed for values larger than $\Delta m$. The peak of the $M_{l^+ l^-}$ distribution shifts to higher energies when the mass gap increases. The invariant mass of the OSSF lepton pair from on shell $Z$ boson decays peaks at $M_Z$.\\
As long as $\Delta m < M_Z$, we can reject events with large ($>60$ GeV) $M_{l^+l^-}$. This makes sure a large signal acceptance remains, while a large fraction of the background arising from on shell $Z$ decays is rejected. We also need to introduce a lower cut on the invariant mass distribution, as the background has an accumulation of events at low $M_{l^+l^-}$ due to events with $J/\psi$ mesons, $\Upsilon$ mesons and low-mass Drell-Yan processes. Usually this cut is set at $M_{l^+ l^-} > 12$ GeV, and we will do the same in this analysis. \\
The selection of the OSSF lepton pair becomes complicated when there is more than one possible OSSF lepton pair. In the ATLAS search, the lepton pair with $M_{l^+ l^-}$ closest to $M_Z$ is taken. We select the OSSF lepton pair as the lepton pair that has a minimal distance in $\Delta R = \sqrt{\Delta \eta^2 +\Delta \phi^2}$ (where $\eta$ indicates the pseudorapidity and $\phi$ the azimuth angle). Because correctly selected pairs have an invariant mass that is kinematically limited, this choice ensures that we have a clearer edge at $\Delta m$, while wrongly selected pairs will usually have a mass that is closer to $M_Z$. \\

Although the LSPs carry away most of the energy of the NLSPs, this does not necessarily mean that there is a large missing energy ($\slashed{E}_T$). In figure~\ref{fig:metb} we see that $\slashed{E}_T$ is, counterintuitively, lower than the $\slashed{E}_T$ originating from the $WZ$ background process. This is due to the fact that the two LSPs are often produced back-to-back. With increasing mass gaps, the final state leptons get more energy, causing them to recoil against the LSP. This causes the LSPs to be produced less back-to-back, which results in a higher missing transverse energy (figure~\ref{fig:diffmassmetb}). 

\clearpage

\subsection{Reducible backgrounds}
\label{sec:finalseleccuts}

\begin{figure}[h]
\vspace{-2cm}
\centering
 \subfigure[]{\label{fig:ptlback}\includegraphics[width=0.48\textwidth]{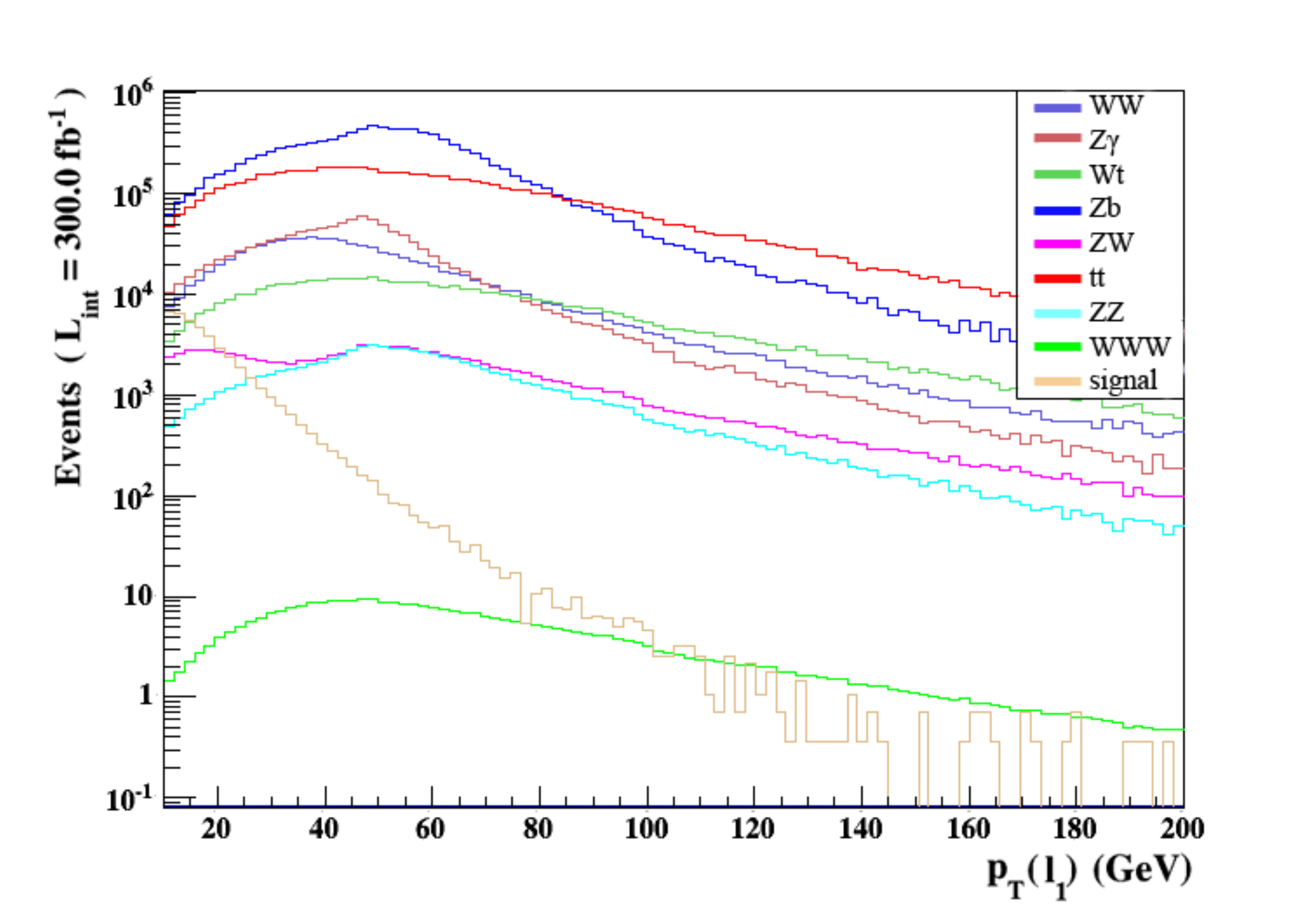}}
 \subfigure[]{\label{fig:ptjback}\includegraphics[width=0.49\textwidth]{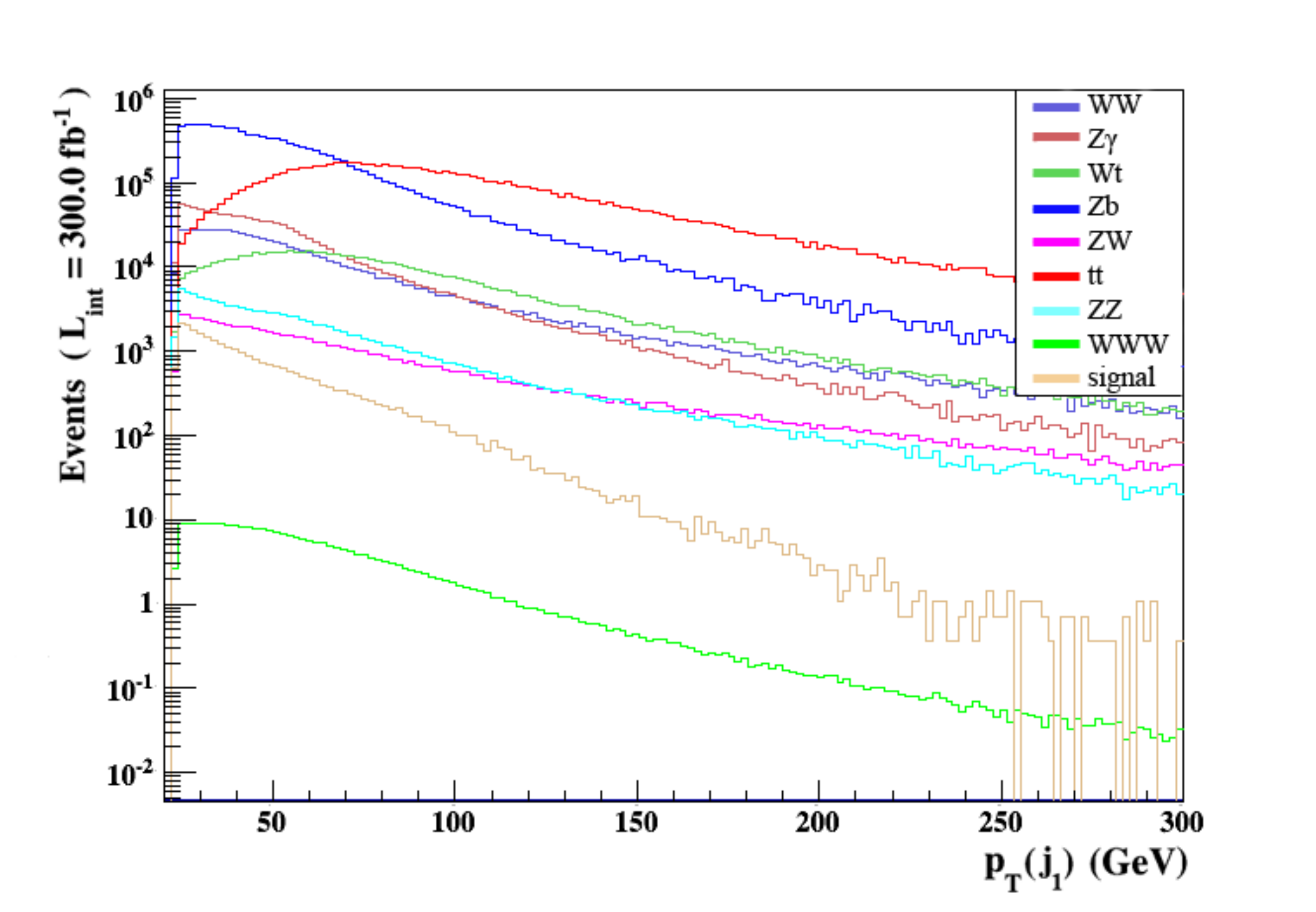}}
\vspace{-0.3cm}
\caption{Distribution for $p_T(l_1)$ (\emph{left}) and $p_T(j_1)$ (\emph{right}) after detector simulation and before imposing any cuts. Shown are all reducible and irreducible background distributions, as well as the distributions we would expect for a $\Delta m = 20$ GeV wino NLSP model with $m_{\tilde{\chi}^0_1} \sim 100$ GeV. Events are normalized to a luminosity of 300 fb$^{-1}$.}

\vspace{-0.15cm}
\centering
 \subfigure[]{\label{fig:backgroundcompa}\includegraphics[width=0.49\textwidth]{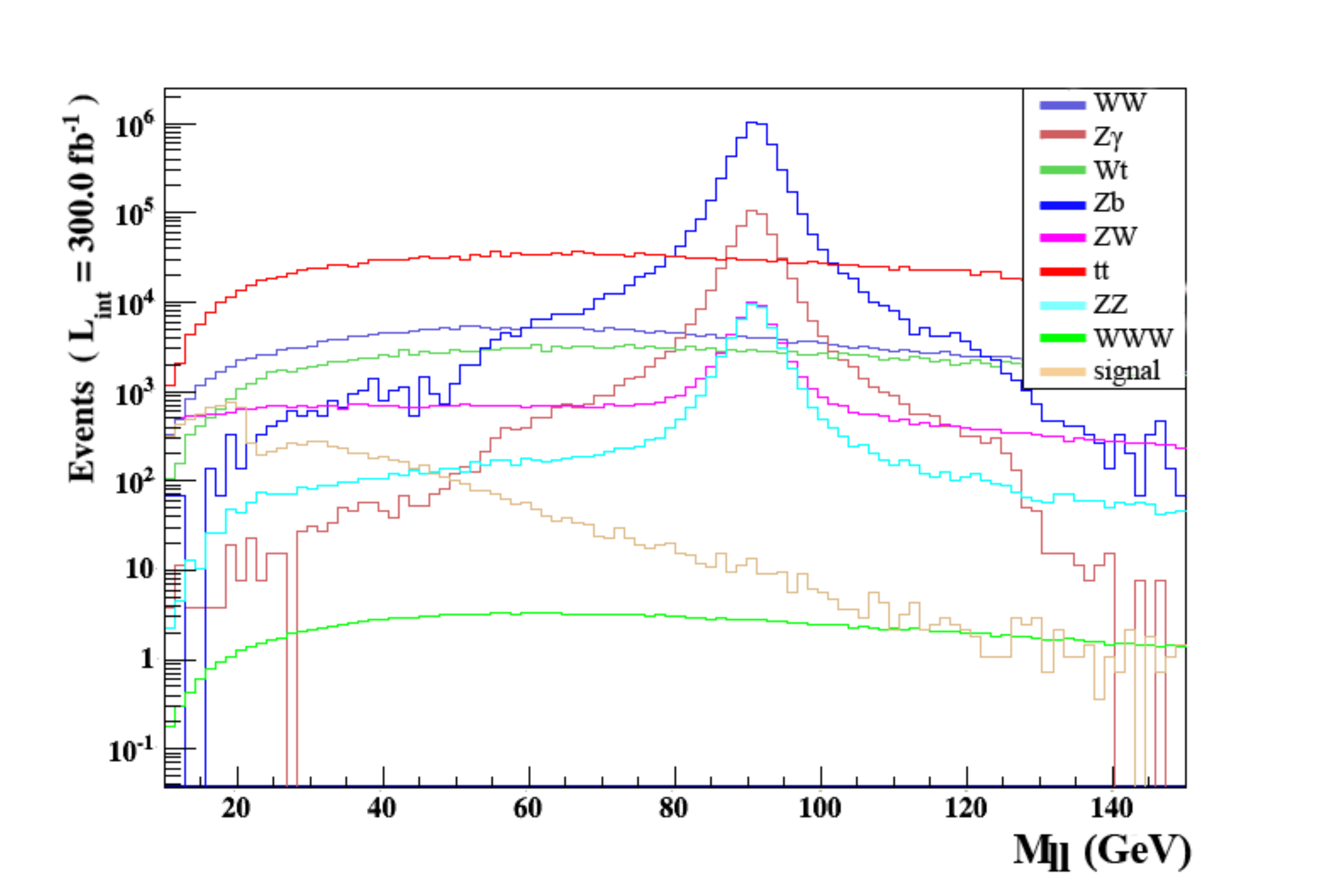}}
 \subfigure[]{\label{fig:backgroundcompb}\includegraphics[width=0.48\textwidth]{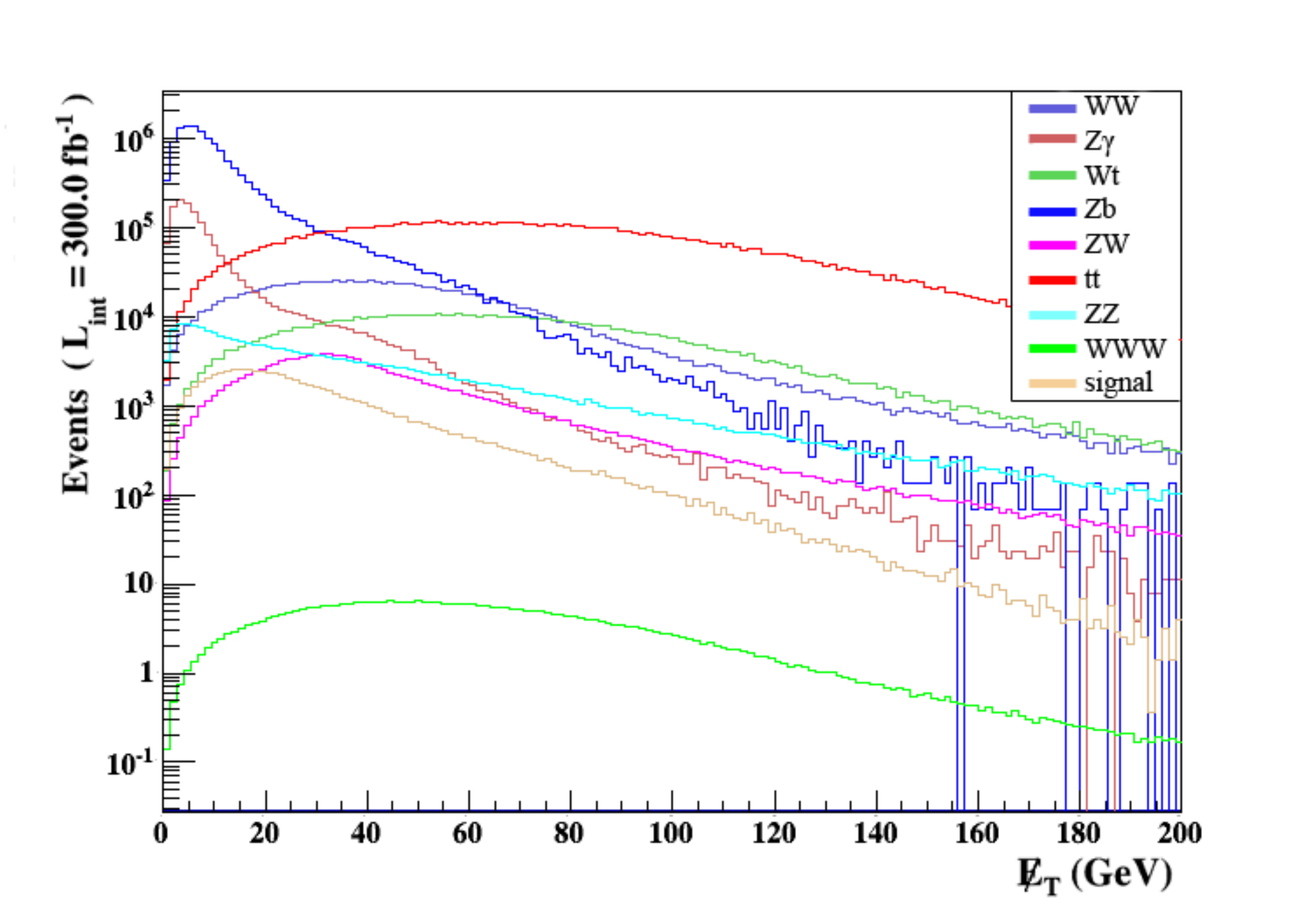}}
\vspace{-0.3cm}
\caption{Distribution for $M_{l^+ l^-}$ (\emph{left}) and $\slashed{E}_T$ (\emph{right}) after detector simulation and before imposing any cuts. Shown are all reducible and irreducible background distributions, as well as the distributions we would expect for a $\Delta m = 20$ GeV wino NLSP model with $m_{\tilde{\chi}^0_1} \sim 100$ GeV. Events are normalized to a luminosity of 300 fb$^{-1}$.}

\vspace{-0.1cm}
 \subfigure[]{\label{fig:metlw}\includegraphics[width=0.47\textwidth]{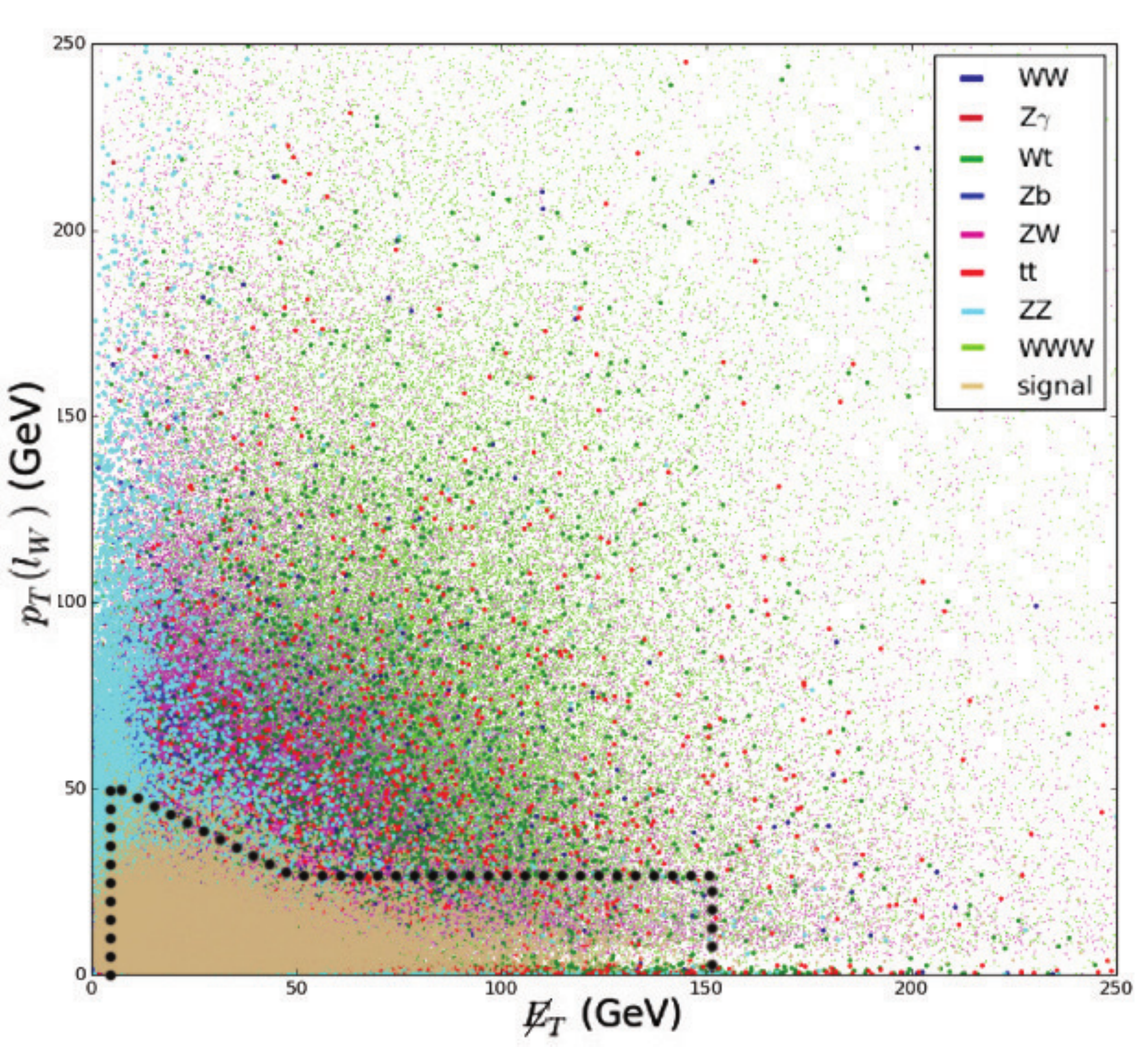}}	
\subfigure[]{\label{fig:delta}\includegraphics[width=0.48\textwidth]{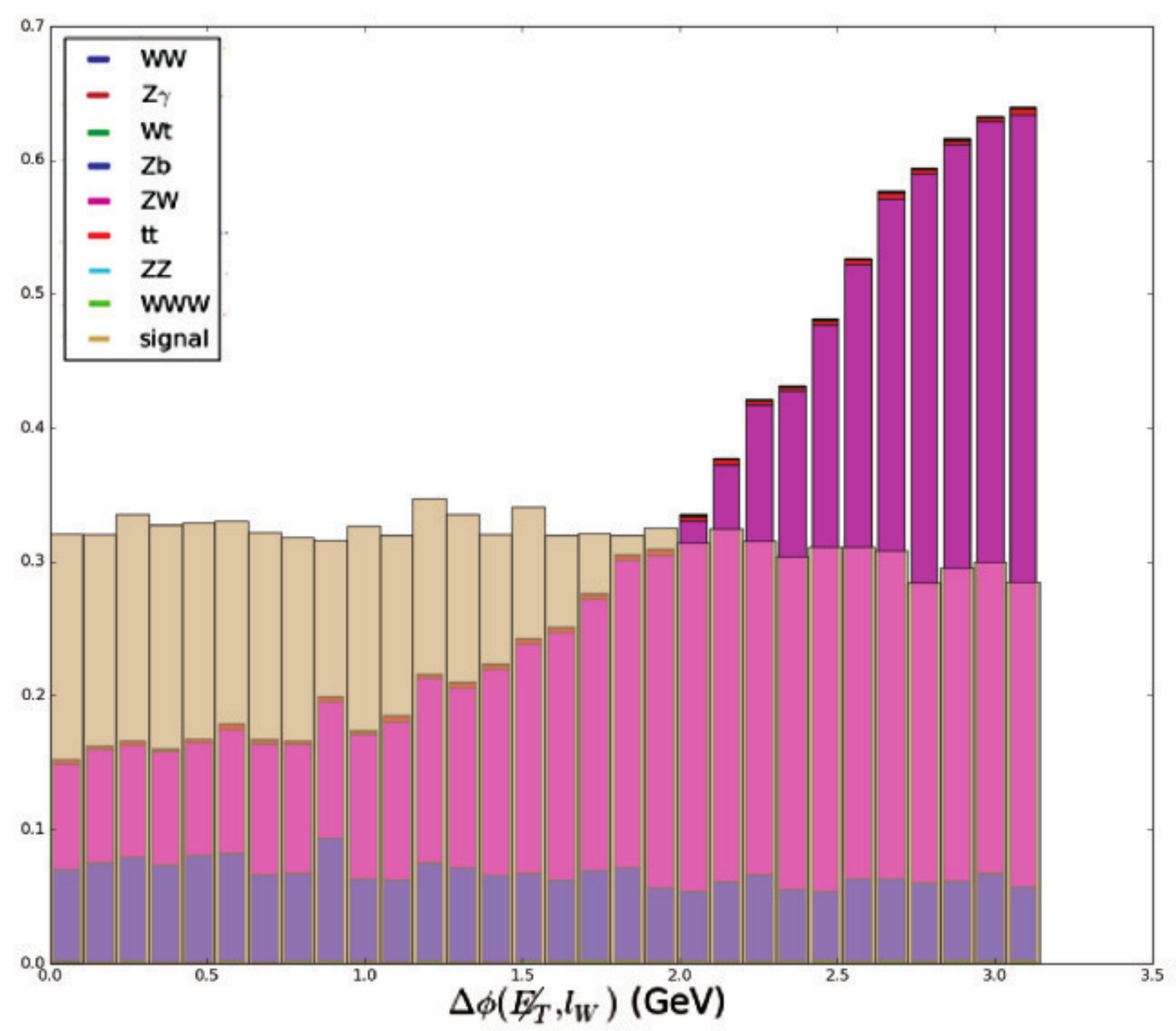}}
\vspace{-0.3cm}
\caption{\emph{Left:} Scatter plot of $\slashed{E}_T$ against $p_T(l_W)$. The dotted black line indicates the \emph{funnel cut} (defined in section \ref{sec:finalseleccuts}). \emph{Right:} Distribution for $\Delta\phi(\slashed{E}_T, l_W)$, the number of events are weighted by their cross section and the histogram is normalized to 1 (for background and signal separately). 
Both figures are made after demanding exactly 3 leptons and at least one OSSF lepton pair. We show the distributions we would expect for a $\Delta m = 20$ GeV wino NLSP model with $m_{\tilde{\chi}^0_1}~\sim~100$~GeV. }
\vspace{-1cm}
\end{figure}

To get rid of a large fraction of the $t\bar{t}$ background, a jet veto with $p_T(j) > 50$ GeV or $p_T(j) > 30$ GeV can be introduced (figure~\ref{fig:ptjback}). The only significant remaining background is then $Zb$, which can be rejected efficiently by the requirement on $M_{l^+ l^-}$, as shown in figure~\ref{fig:backgroundcompa}.\\
The transverse momentum of the highest $p_T$ lepton ($p_T(l_1)$) will be smaller for the signal than for the background, as shown in figure~\ref{fig:ptlback}. We therefore can also veto events with high $p_T(l)$. As can be seen in figure~\ref{fig:backgroundcompb}, the $\slashed{E}_T$ of reducible backgrounds will be mostly larger than the one of the signal. 
We expect a higher $\slashed{E}_T$ and $p_T(l_W)$ (transverse momentum of the lepton originating from $W$-decay) for background events than for the signal events (figure~\ref{fig:metlw}). The signal events for low $p_T(l_W)$ and $\slashed{E}_T$ are correlated in a funnel-like shape. We will use this feature to discriminate signal from background. 

In the rest frame of a particle decaying in two other particles, the decay particles will be produced back-to-back. We therefore expect that the background distribution for $\Delta\phi(\slashed{E}_T, l_W)$ is peaked towards $\Delta\phi = \pi$ for any events containing $\slashed{E}_T$ and $l_W$ originating from a $W$ boson decay. If we then allow for a boost of the $W$ bosons, the distribution will get smeared out to other values as well, although a small peak at $\Delta\phi = \pi$ remains. We do not expect the same topology for the signal events. This is because the $\slashed{E}_T$ will now be the sum of three components: two LSPs and a neutrino. We expect that $\Delta\phi(\slashed{E}_T, l_W)$ will be uniformly distributed for the signal events. This is also observed in the Monte Carlo generated events, as shown in figure~\ref{fig:delta}. \\

To conclude, we now have 5 observables that we can use to discriminate signal from background:
\begin{equation*}
M(l^+ l^-), \hspace{0.5cm} p_T(l), \hspace{0.5cm} p_T(j), \hspace{0.5cm} \slashed{E}_T \text{ vs } p_T(l_W), \hspace{0.5cm} \Delta\phi(\slashed{E}_T, l_W).
\end{equation*}

Based on the features just discussed, we use the following cuts to optimize the analysis:
\begin{itemize}

\item $N(l) = 3$ and $N(l^+ l^-) > 0$ (at least one OSSF lepton pair).
\item $5 $ ${\rm GeV} < p_T(\mu) < 50$ GeV and $10$ $ {\rm GeV} < p_T(e) < 50$ GeV. 
\item $12$ $ {\rm GeV} < M(l^+ l^-) < 60$ GeV.
\item Veto on jets with $p_T(j_1) > 50$ or 30 GeV and $|\eta(j_1)| < 2.5$.
\item \emph{Funnel cut:} (dotted black line in figure~\ref{fig:metlw}) $5$ $ {\rm GeV} <\slashed{E}_T < 150$ GeV \textbf{and} 
     \begin{itemize}
	\item if ${\slashed{E}_T} < 50$ GeV: $p_T(l_W) + 0.6{\slashed{E}_T} < 50$ GeV
	\item else: $p_T(l_W) < 20$ GeV. 
     \end{itemize}
\item $\Delta\phi(\slashed{E}_T, l_W) < 2$.
\end{itemize}
The cut flow diagram for the background processes and the four benchmark SUSY scenarios is given in table~\ref{tab:cutflow}. 

\begin{sidewaystable}[h]
\centering
\begin{tabular}{ |l|c|c|c|c|c|c|c|c|}
\hline
& $WZ$ & $WW$ & $ZZ$ & $Zb$ & $Wt$ & $Z\gamma$ & $t\bar{t}$ & $WWW$ \\
\hline
\hline
Before cuts &132705&1244002&248389&13606295&716358&1453092&8736325&402\\ 
\hline
$N(l) = 3$&60357&3450&5099&415519&16970&4442&83556&120\\ 
$N(l^+ l^-) > 0$ &60300&3439&5095&413974&16933&4431&83118&120\\ 
$5 $ GeV $< p_T(\mu) < 50$ GeV and $10$ GeV $< p_T(e) < 50$ GeV &19323&1470&863&139715&4349&1987&22816&23\\
$12$ GeV $ < M(l^+ l^-) < 60$ GeV&8573&711&195&25592&2528&233&12218&13\\ 
$p_T(j) < 50$ GeV  &7638&599&136&23039&974&218&2496&12\\
Funnel cut&5067&426&35&10881&572&61&1269&4\\ 
$\Delta\phi(\slashed{E}_T, l_W) < 2$ &3327&354&8&2955&467&7&1051&1\\ 
$p_T(j_1) < 30$ GeV &2979&278&5&1343&176&7&306&1\\ 
\hline
\hline
& \multicolumn{6}{c}{Total background after cuts:} & \multicolumn{2}{c|}{8170 (5095)} \\
\hline
\end{tabular}
\caption{Expected number of events after each consecutive cut at 300 fb$^{-1}$ for the background processes. Statistical uncertainties are not included here. The total expected number of background events assuming $p_T(j) < 50 $ GeV is indicated in the last row, the total expected number of background events assuming $p_T(j) < 30$ GeV is in parentheses.} 
\vspace{0.5cm}
\label{tab:cutflow}
\begin{tabular}{|l| c|c|c|c|}
\hline
& Wino$_{\Delta m = 20 {\rm GeV}}$ &  Wino$_{\Delta m = 50 {\rm GeV}}$ & Wino$_{\Delta m = 100 {\rm GeV}}$ & Higgsino$_{\Delta m = 20 {\rm GeV}}$ \\
\hline \hline
Before cuts&71700&87600&22380&23790\\ 
\hline
$N(l) = 3$&17819&24613&7826&5407\\ 
$N(l^+ l^-) > 0$&17792&24576&7818&5399\\ 
$5 $ GeV $ < p_T(\mu) < 50$ GeV and $10$ GeV $ < p_T(e) < 50$ GeV &12365&17228&971&3889\\ 
$12$ GeV $ < M(l^+ l^-) < 60$ GeV&10018&16597&219&2661\\ 
$p_T(j) < 50$ GeV &8626&14725&187&2310\\ 
Funnel cut&7983&10964&41&2128\\ 
$\Delta\phi(\slashed{E}_T, l_W) < 2$ &5424&6624&26&1397\\ 
$p_T(j_1) < 30$ GeV &4516&5650&19&1166\\ 
\hline 
\end{tabular}
\caption{Expected number of signal events after each consecutive cut at 300 fb$^{-1}$ for four scenarios: wino-like NLSP with different mass gaps (20, 50 and 100 GeV), and higgsino-like NLSP with $\Delta m = 20$ GeV.}
\end{sidewaystable} 
\clearpage

\section{LHC14 reach}
\label{sec:res}

\begin{figure}[h]
\centering
 \subfigure[]{\label{fig:winoatl}\includegraphics[width=0.48\textwidth]{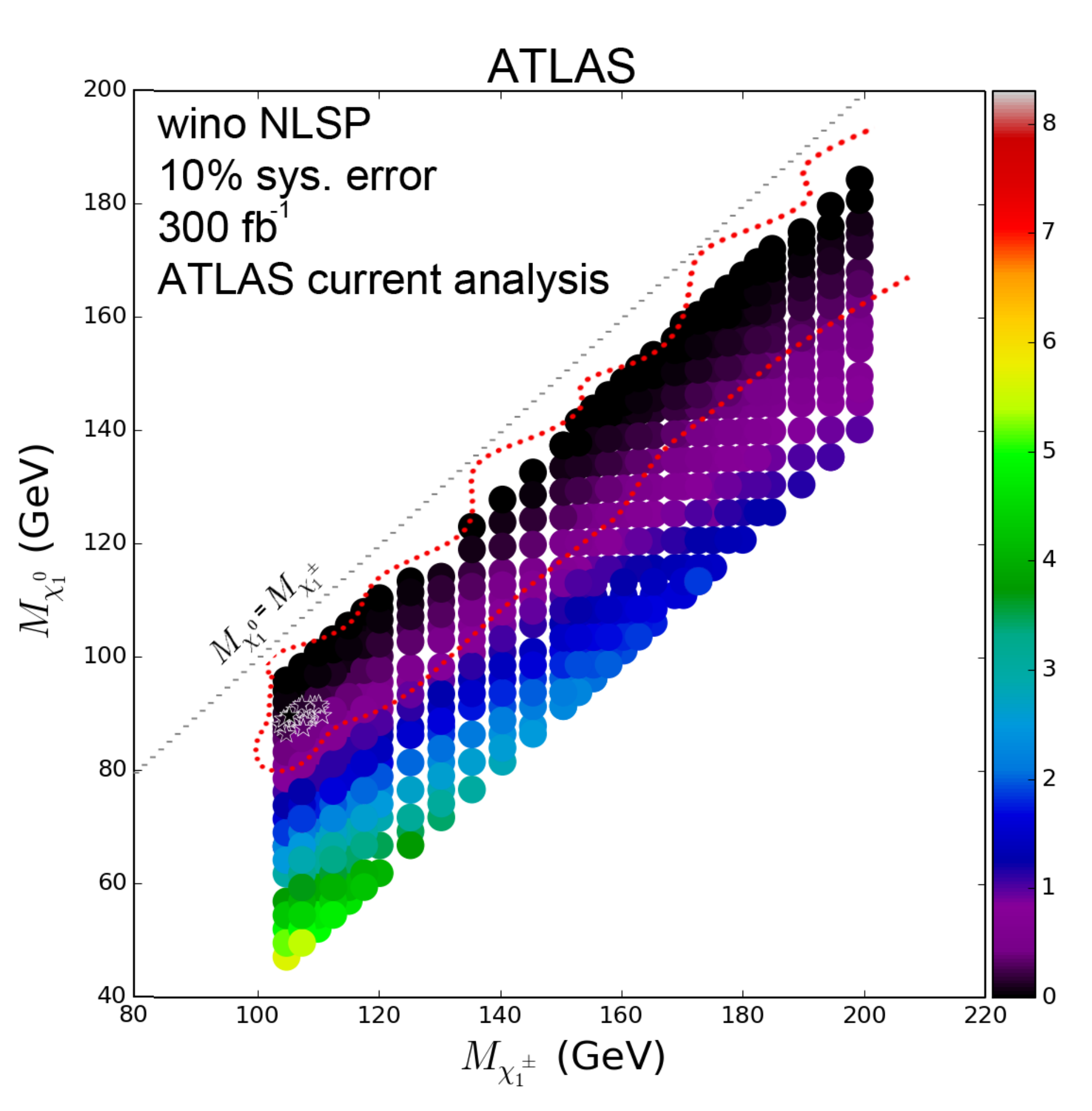}}
 \subfigure[]{\label{fig:hinoatl}\includegraphics[width=0.48\textwidth]{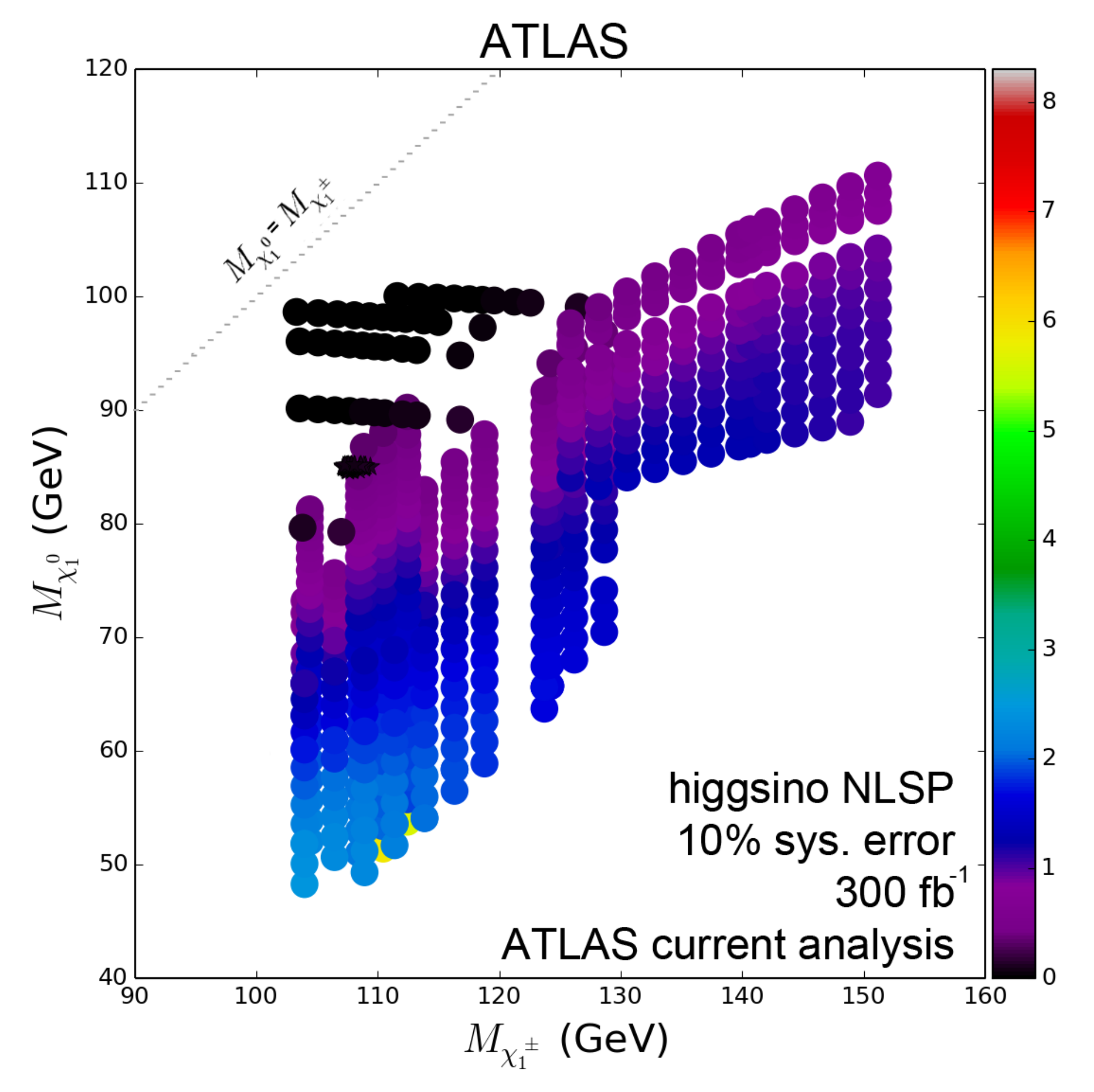}}\vspace{-0.4cm}
\subfigure[]{\label{fig:winocms}\includegraphics[width=0.48\textwidth]{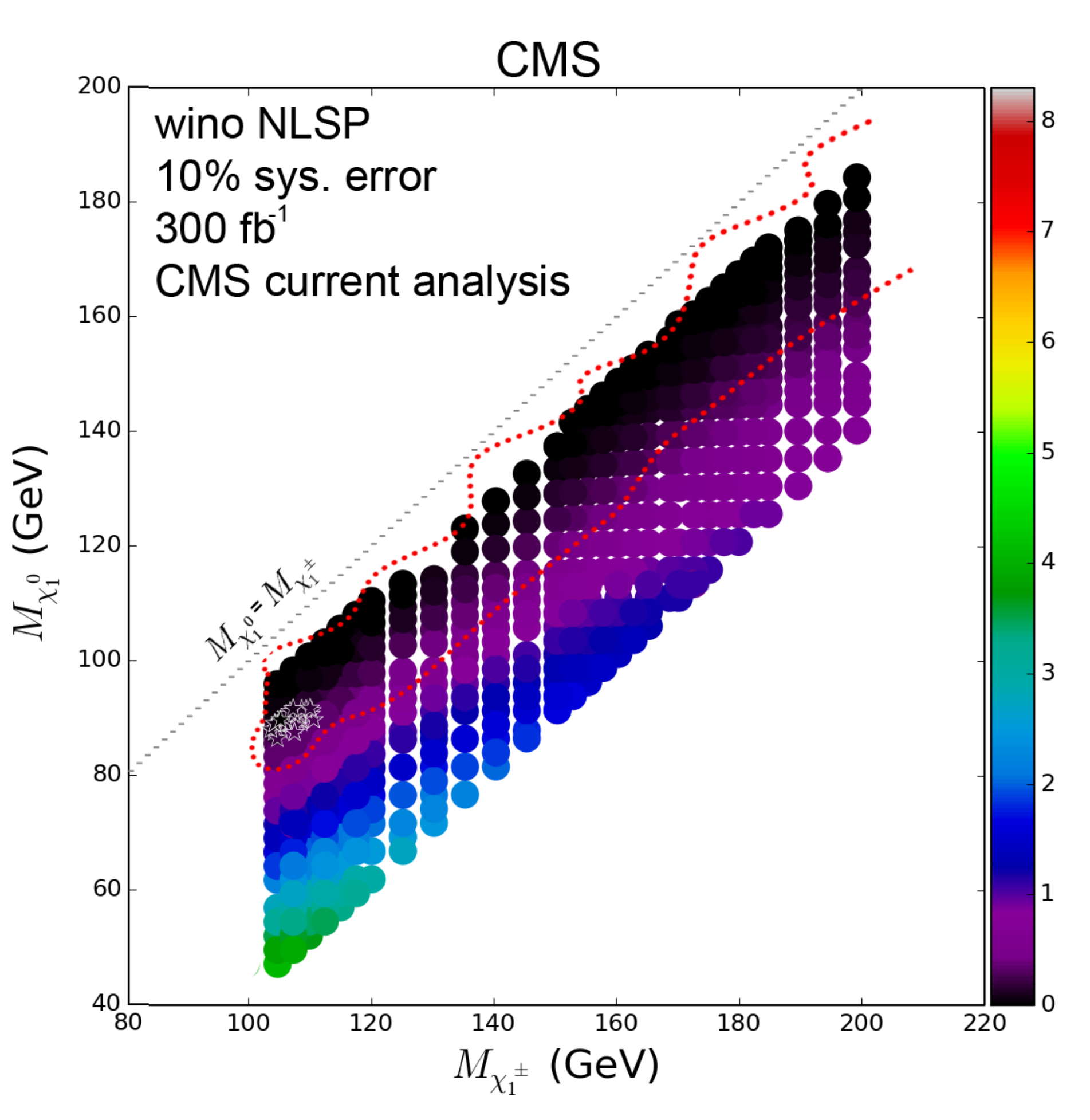}}\vspace{-0.3cm}
 \subfigure[]{\label{fig:hinocms}\includegraphics[width=0.48\textwidth]{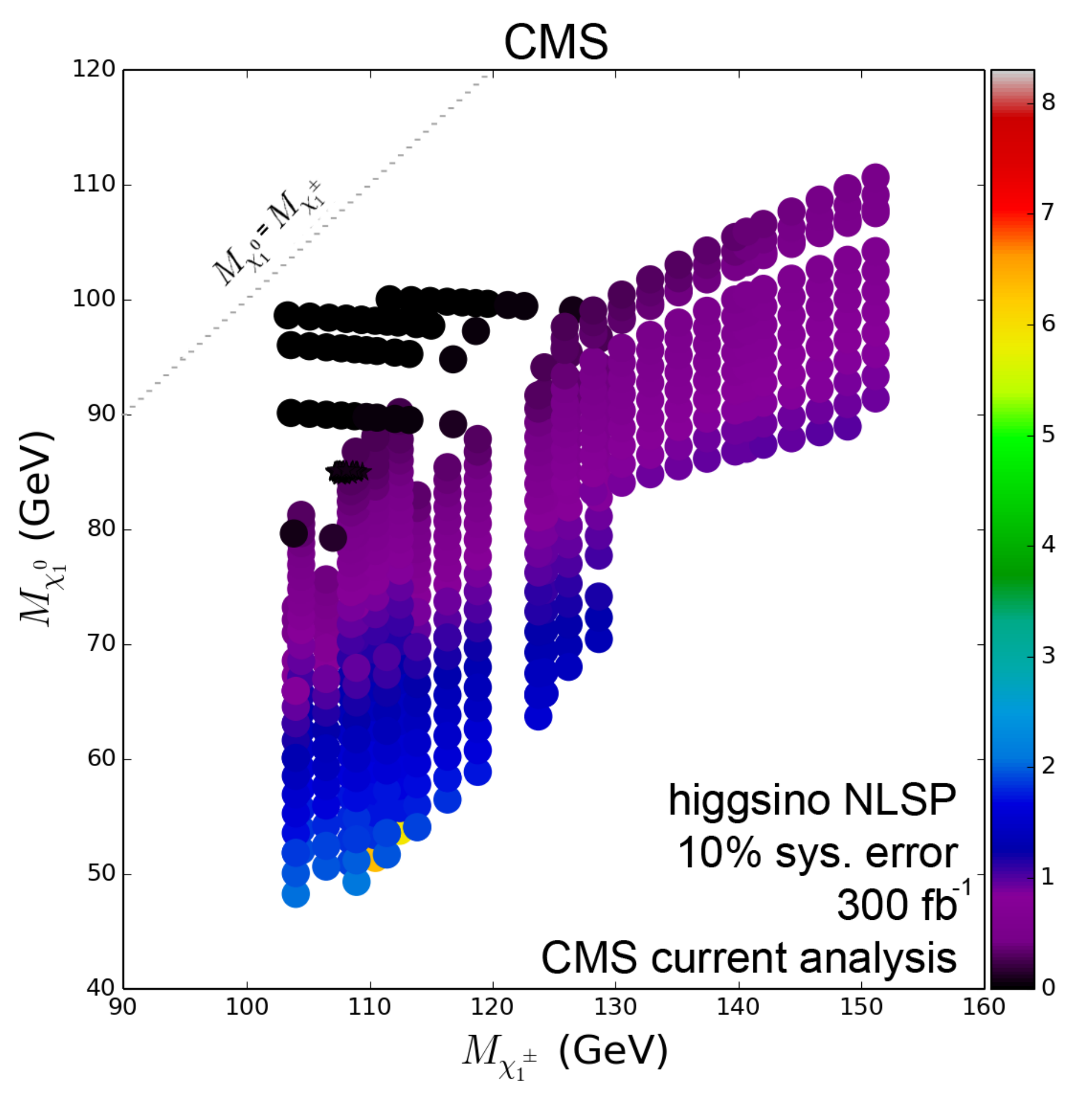}}
\caption{Standard ATLAS and CMS tri-lepton search sensitivities for wino NLSP (figures a and c), and higgsino NLSP (figures b and d) pMSSM models assuming 300 fb$^{-1}$ and 10$\%$ systematic error. Note that the significance (indicated by the color scale) is somewhat lower than for the simplified SUSY models (as indicated in figure~\ref{fig:EWlim}). This is due to the  reduced cross sections and branching ratios in comparison with the simplified SUSY models. The dashed gray line indicates the limit $m_{\tilde{\chi}^0_1} = m_{\tilde{\chi}^{\pm}_1}$. Stars (located around $m_{\tilde{\chi}^{\pm}_1} = 110$ GeV) indicate the GC best fit pMSSM models from ref.~\cite{Caron:2015wda}, which coincide with the best global fit models obtained by~\cite{2015arXiv150707008B}.  The dotted red line indicates the $1\sigma$ contour of the most likely pMSSM10 models from ref.~\cite{Bagnaschi:2015eha} (only for wino NLSP). }
\end{figure}

To evaluate our sensitivity, we will use the $Z_N$ value (as defined in ref.~\cite{2003sppp.conf...35L}) as the significance and assume a systematic background error of $\sigma = 10 \%$. The $Z_N$ value measures the difference between the outcome of a pMSSM model and the outcome of the Standard Model in units of the standard deviation. Typically, the exclusion reach is indicated by the 95$\%$ CL, which corresponds to a significance of $2\sigma$ (and therefore a $Z_N$ value of 2). In figure~\ref{fig:winoatl} and~\ref{fig:winocms} (\ref{fig:hinoatl} and \ref{fig:hinocms}) we present the significance as a color code for the ATLAS and CMS reach for the wino (higgsino) NLSP scenarios using their current tri-lepton searches (as indicated in table~\ref{tab:searches}).  
The dotted red line indicates the $1\sigma$ contour of the most likely pMSSM10 models from ref.~\cite{Bagnaschi:2015eha}, which was found for wino NLSP models. The higgsino NLSP production cross section is smaller than the wino NLSP production cross section, which reduces the significance. We observe that the sensitivity for ATLAS and CMS increases for higher $\Delta m$. The sensitivity using the ATLAS search strategy reaches $>2\sigma$ for mass gaps $>30$ GeV and LSP masses $<100$ GeV for the wino NLSP pMSSM models. Note that this does not exactly resemble the ATLAS limit indicated by the purple line in figure~\ref{fig:EWlim}. This is because, in contrast with ATLAS, we do not use simplified models where the NLSPs are $100\%$ wino, which reduces the neutralino-chargino production cross section by a factor $\sim$1.15. Furthermore, the branching ratios for the $\tilde{\chi}^0_2$ and $\tilde{\chi}^{\pm}_1$ decays to the $Z$ and $W$ boson are set at 100$\%$ in the simplified models, whereas in our models these branching ratios are not 100$\%$.\\

\begin{figure}[h]
\centering
 \subfigure[]{\label{fig:winopt5}\includegraphics[width=0.49\textwidth]{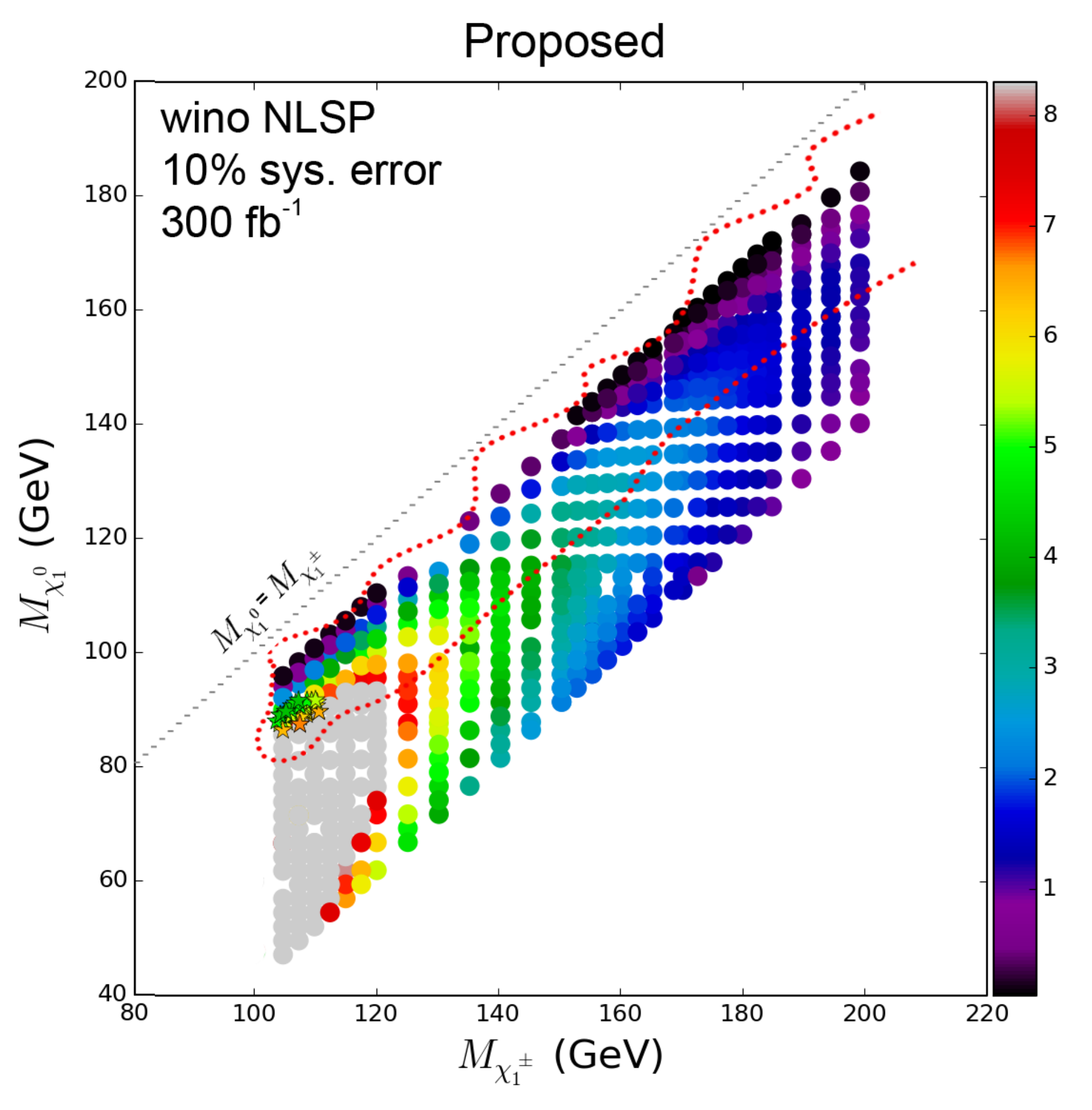}}
 \subfigure[]{\label{fig:higgsinopt5}\includegraphics[width=0.49\textwidth]{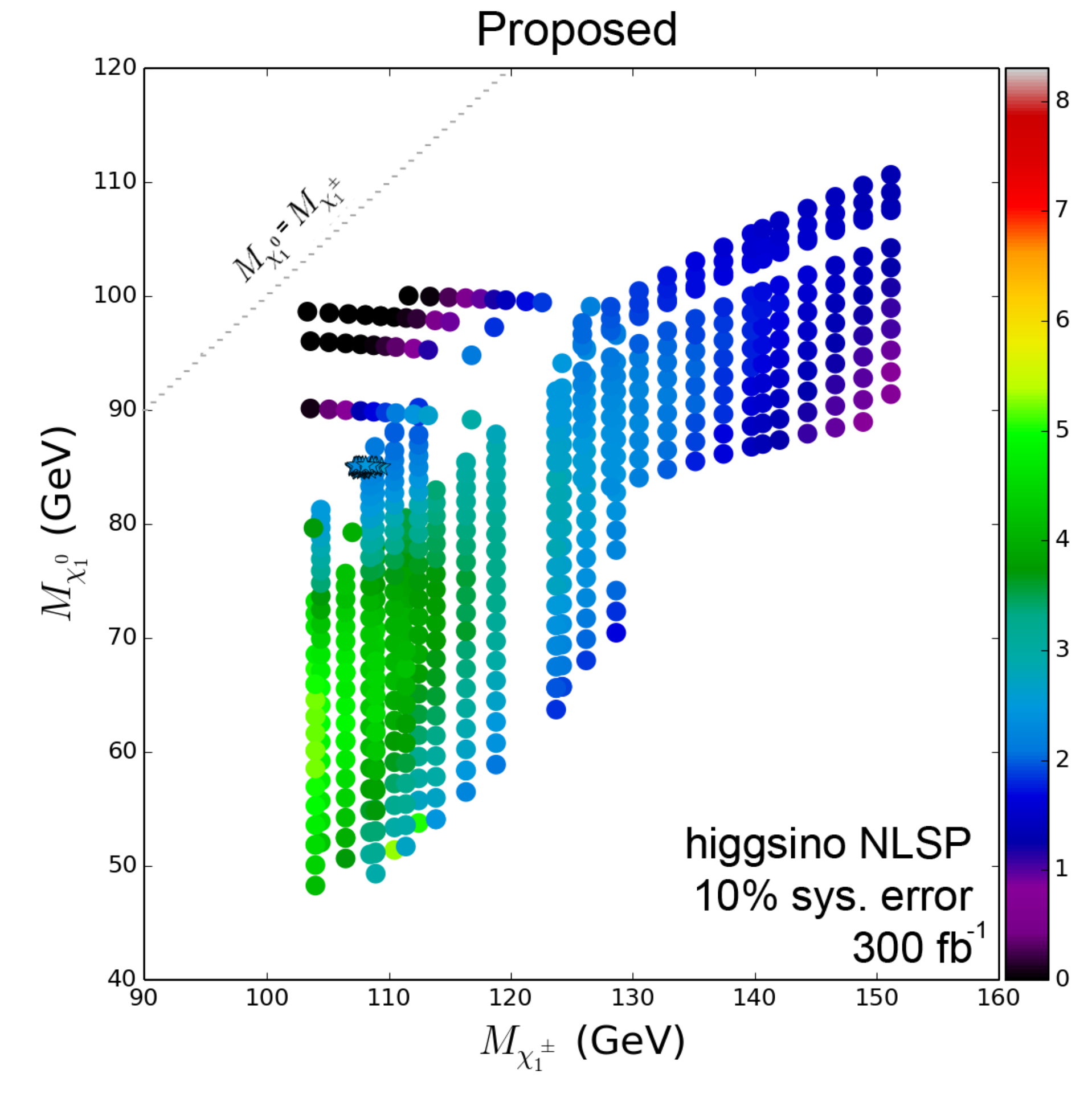}}
\caption{ Significance (indicated by the color scale) for wino NLSP (a) and higgsino NLSP (b) pMSSM models assuming a background systematic uncertainty of $10\%$. The dashed gray line indicates the limit $m_{\tilde{\chi}^0_1} = m_{\tilde{\chi}^{\pm}_1}$. Stars (located around $m_{\tilde{\chi}^{\pm}_1} = 110$ GeV) indicate the GC best fit pMSSM models from ref.~\cite{Caron:2015wda}, which coincide with the best global fit models obtained by~\cite{2015arXiv150707008B}.  The dotted red line indicates the $1\sigma$ contour of the most likely pMSSM10 models from ref.~\cite{Bagnaschi:2015eha} (only for wino NLSP).}
\label{fig:limitresults}
\vspace{-0.5cm}
\end{figure}
However, using the proposed search strategy with $10\%$ background uncertainty, the sensitivity would be greatly enhanced compared to the standard LHC searches. A comparison is given in figure~\ref{fig:limitresults} for the wino NLSP and higgsino NLSP scenarios and an integrated luminosity of 300 fb$^{-1}$. We find that the 14 TeV LHC can probe LSP masses up to 140 GeV for mass gaps between $\sim 9-50$ GeV if the NLSPs are wino-like (figure~\ref{fig:winopt5}) and LSP masses up to 95 GeV for $\Delta m \gtrsim 20$ GeV if the NLSPs are higgsino-like (figure~\ref{fig:higgsinopt5}), using a jet veto with $p_T(j) < 30$ GeV. The reduced production cross section is the limiting factor for higher LSP and NLSP masses for both pMSSM scenarios. Evidently, these special small-mass-gap SUSY scenarios for heavy sfermions have large repercussions on the LHC SUSY search strategy. Studies suggest that these models are most likely to be realized in nature, but we cannot rely on the standard high jet $p_T$ or $\slashed{E}_T$ triggers for its discovery. In contrast, using the proposed search strategy it is even possible to probe the discussed SUSY scenarios at an integrated luminosity of 30 fb$^{-1}$. A detailed discussion is presented in the appendix.
\clearpage

\section{Conclusion} 
Global pMSSM fits and the pMSSM solution of the GC photon excess suggest a $\sim$100 GeV bino-like lightest neutralino as a viable WIMP candidate, accompanied by a chargino and neutralino that are 10-25 GeV heavier. Standard mono-jet searches are not sensitive to these models due to the large bino component of the lightest neutralino. We found that the current LHC electroweak SUSY search strategies are not and will not be sensitive to these favored pMSSM models. We therefore propose an improved search strategy to enhance the sensitivity of the  $3l + \slashed{E}_T$ final state searches at the LHC for precisely these pMSSM scenarios. The main irreducible background for this search channel is the production of $WZ$, where the bosons decay leptonically. The main reducible background processes are $t\bar{t}$ and $Zb$. Contrary to what is required in most searches for SUSY, we find that the requirement of \emph{low} missing transverse energy increases the sensitivity to these pMSSM models compared to the current electroweak SUSY searches. \\
\begin{figure}[h]

\begin{center}
\subfigure[]{\label{fig:concl1}\includegraphics[width=0.46\textwidth]{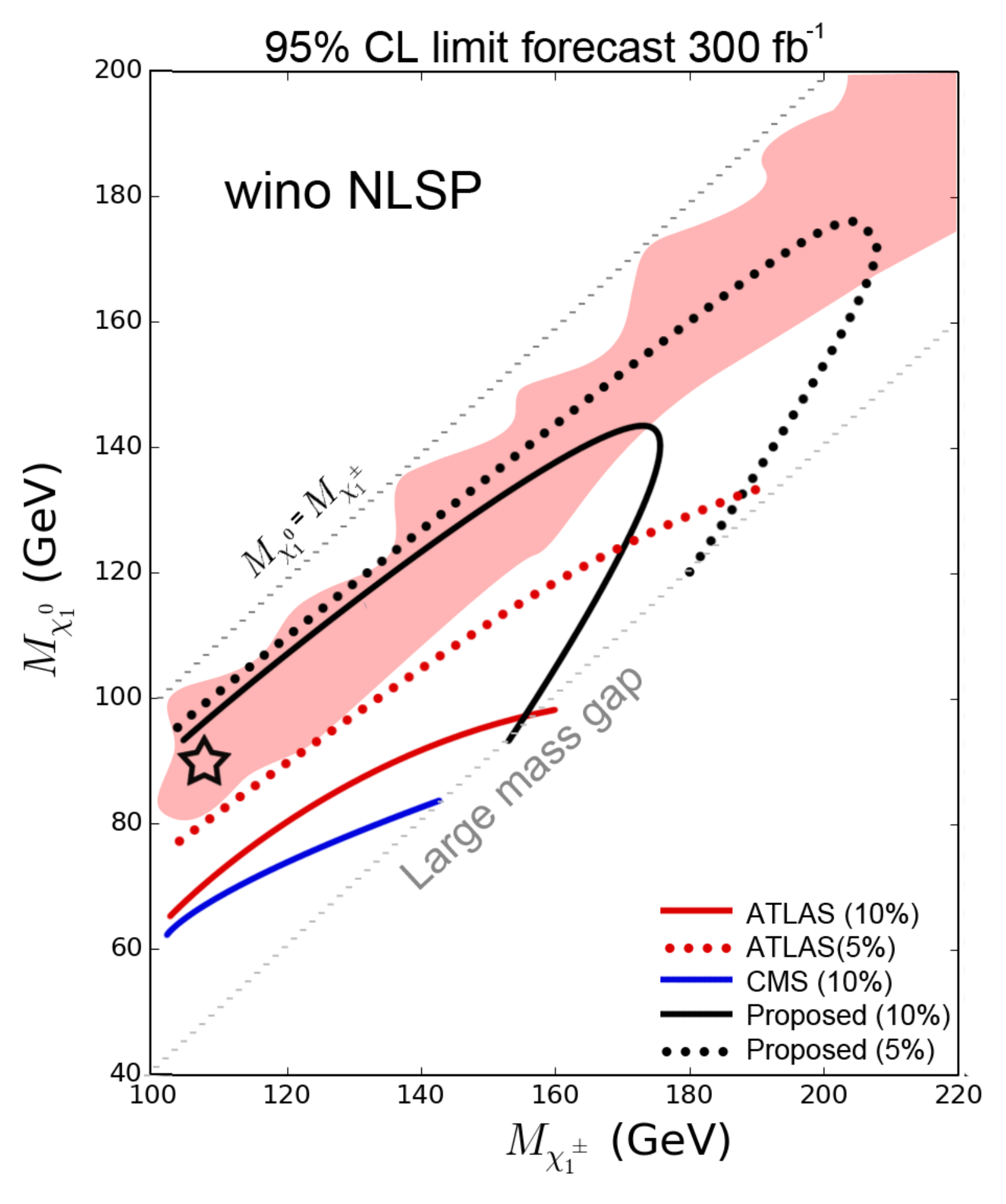}}
\subfigure[]{\label{fig:concl2}\includegraphics[width=0.49\textwidth]{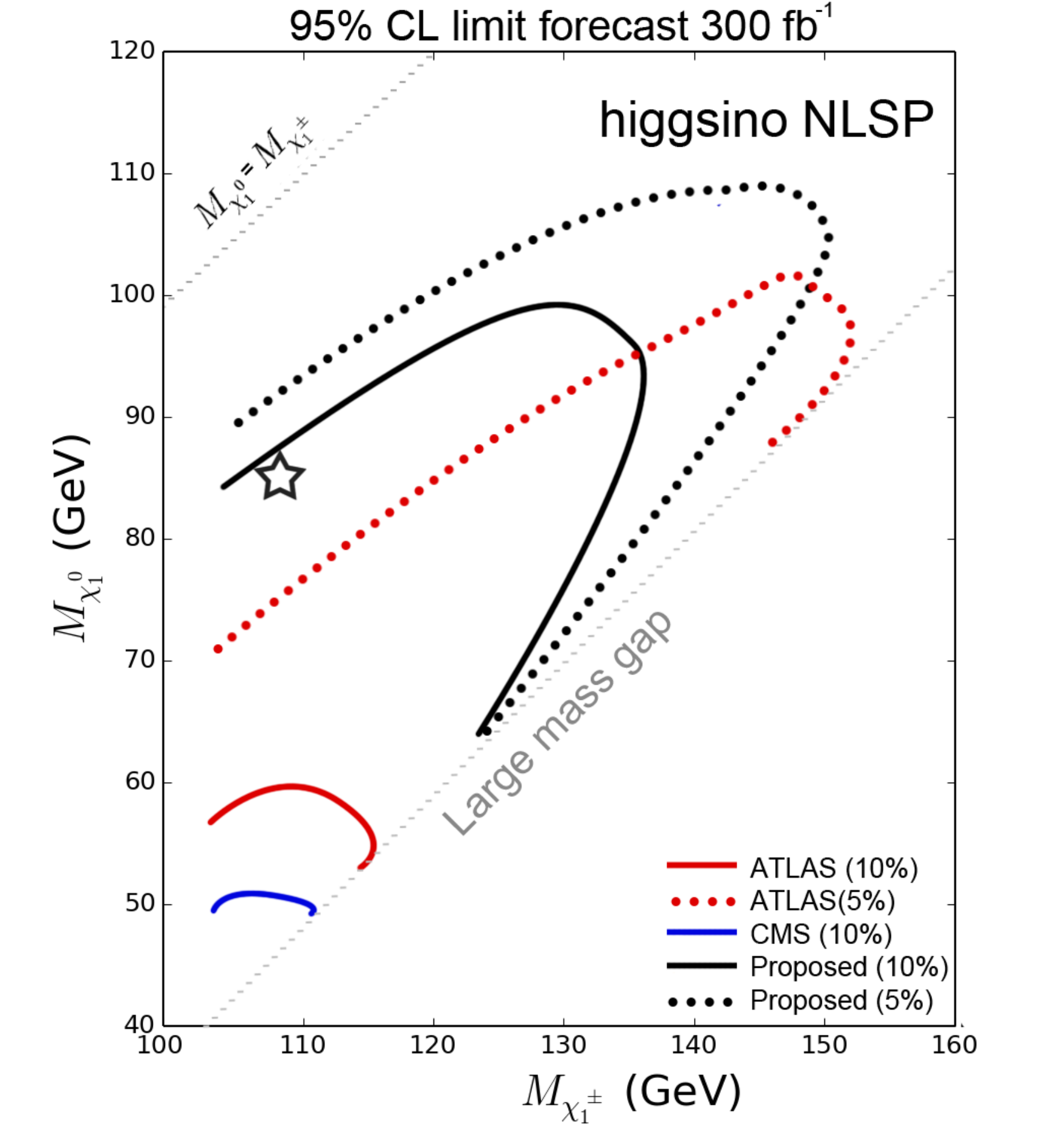}}
\caption{The expected $2\sigma$ exclusion reach for the LHC at 14 TeV and with 300 fb$^{-1}$ of data for the wino NLSP (a) and higgsino NLSP (b) models. The current CMS and ATLAS reach is indicated by the blue and red solid (dotted) line, using a systematic background uncertainty of 10$\%$ (5$\%$). The solid (dotted) black line indicates the limit obtained using the default lepton transverse momentum cuts with the requirement $p_T(j) < 30$ GeV, using a background uncertainty of 10$\%$ (5$\%$). Stars indicate the GC best fit pMSSM models from ref.~\cite{Caron:2015wda}, which coincide with the best global fit points obtained by~\cite{2015arXiv150707008B}.  The shaded red area indicates the $1\sigma$ contour of the most likely pMSSM10 models from ref.~\cite{Bagnaschi:2015eha} (only for wino NLSP).}
		
		\label{fig:conc}
	\end{center}

\end{figure}
\clearpage
Using the proposed strategy, the 14 TeV LHC with 300 fb$^{-1}$ integrated luminosity could probe bino-like DM with masses up to 140 GeV using the chargino-neutralino production channel and could go down to mass splittings as low as 9 GeV. We stress the importance of a dedicated SUSY search that targets compressed pMSSM scenarios with a bino-like LSP, as these pMSSM scenarios are favored by global fit studies and by the photon excess spectrum observed for the Galactic Center. Via the introduction of the \emph{funnel cut} and a cut on the invariant mass of the lepton pair originating from leptonic $Z$ decay, the sensitivity for these scenarios is increased tremendously as compared to the current ATLAS and CMS tri-lepton searches, as shown in figure~\ref{fig:conc}. \\



\paragraph{Acknowledgments.} 
R. RdA, is supported by the Ram\'on y Cajal program of the Spanish MICINN and also thanks the support of the Spanish MICINN's Consolider-Ingenio 2010 Programme 
under the grant MULTIDARK CSD2209-00064, the Invisibles European ITN project (FP7-PEOPLE-2011-ITN, PITN-GA-2011-289442-INVISIBLES and the 
``SOM Sabor y origen de la Materia" (FPA2011-29678) and the ``Fenomenologia y Cosmologia de la Fisica mas alla del Modelo Estandar e lmplicaciones Experimentales 
en la era del LHC" (FPA2010-17747) MEC projects.  \\

\clearpage

\section*{Appendix: Detailed discussion of the LHC14 reach}
\begin{figure}[h]
\centering
\subfigure[]{\label{fig:winoatl50}\includegraphics[width=0.485\textwidth]{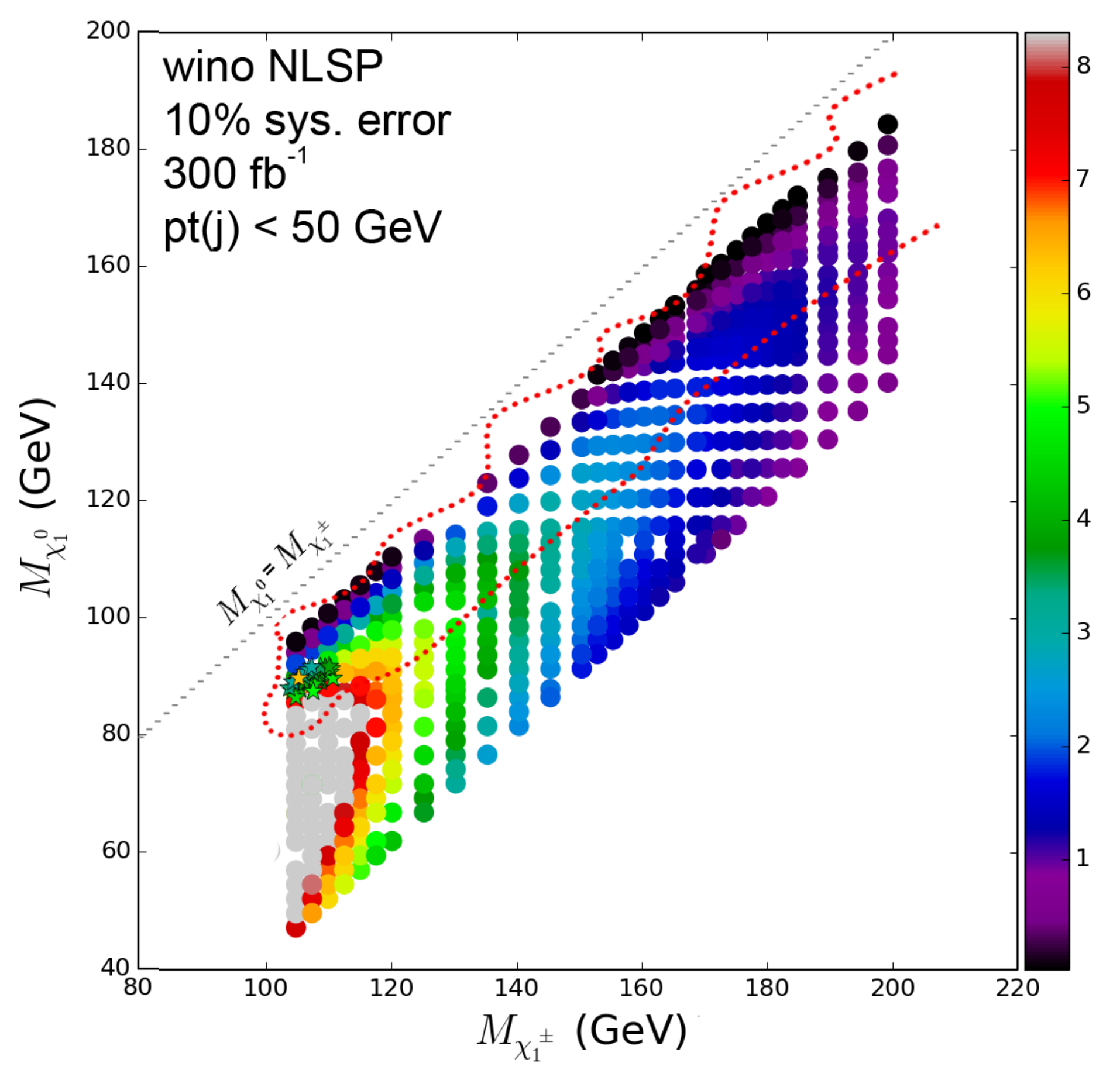}}
\subfigure[]{\label{fig:hinoatl50}\includegraphics[width=0.495\textwidth]{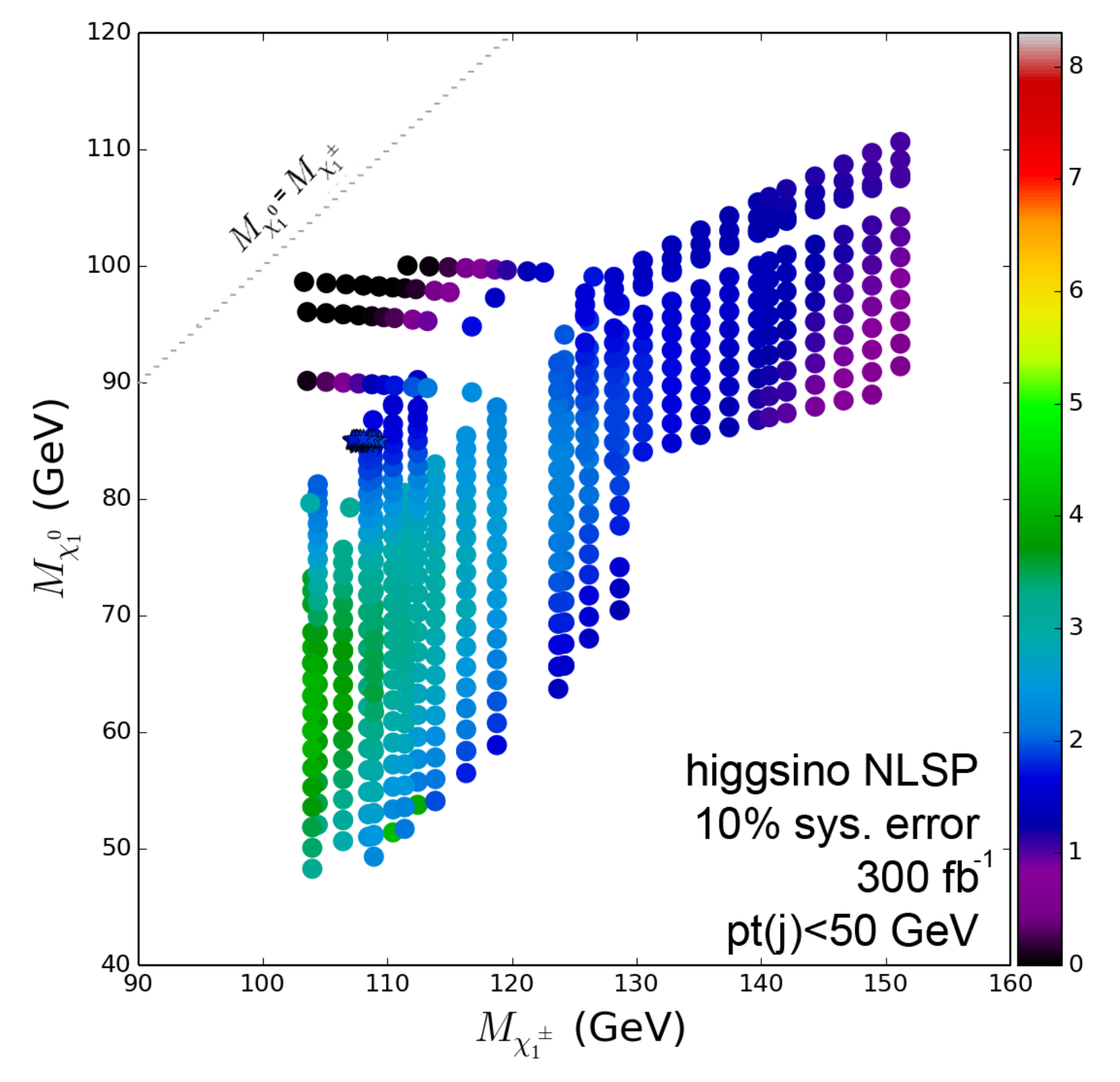}}

\subfigure[]{\label{fig:winoatl30}\includegraphics[width=0.485\textwidth]{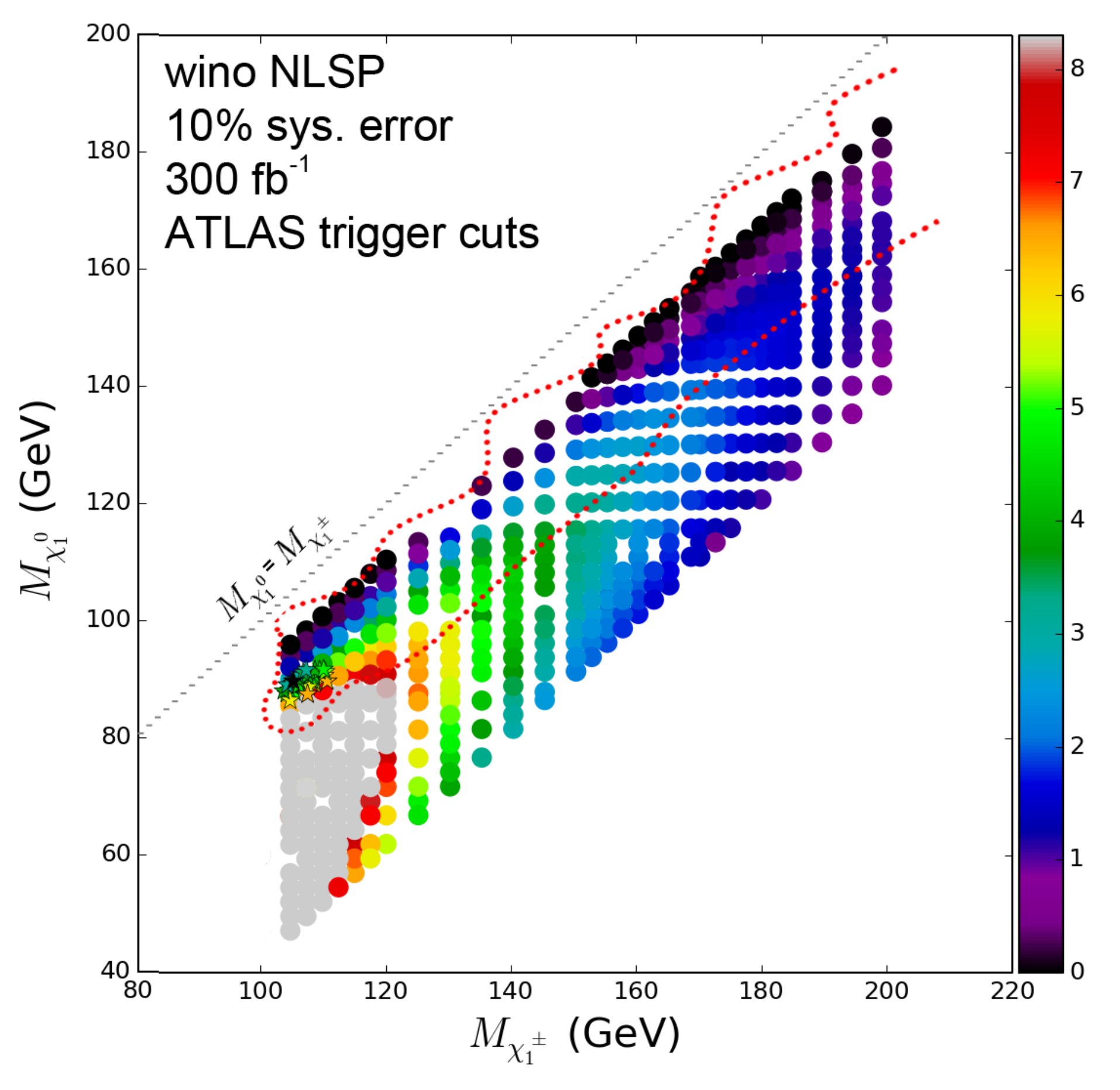}}
\subfigure[]{\label{fig:hinoatl30}\includegraphics[width=0.485\textwidth]{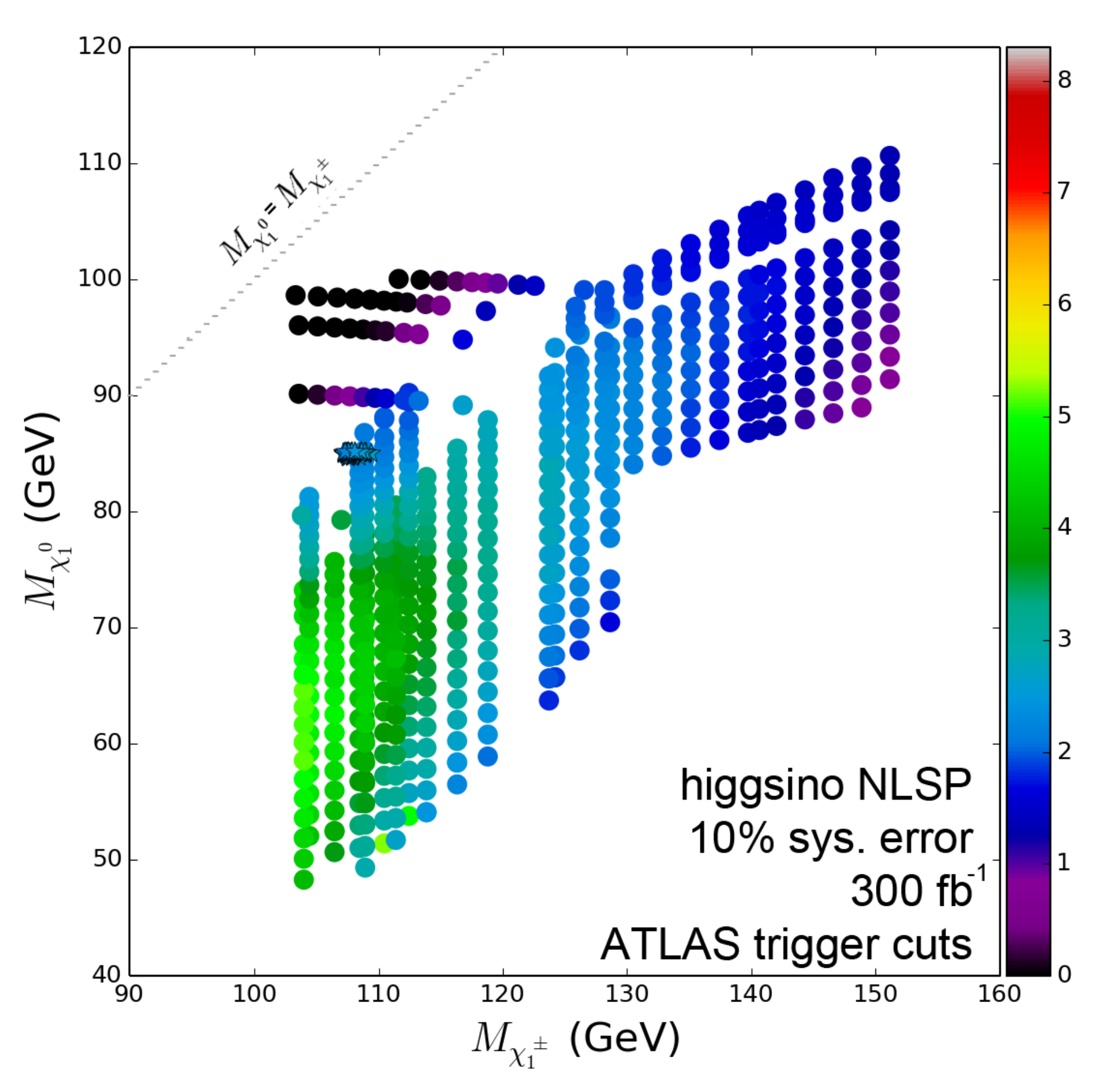}}
\caption{Significance (indicated by the color scale)  for wino NLSP models (\emph{left}) and higgsino NLSP models (\emph{right}). Stars (located around $m_{\tilde{\chi}^{\pm}_1} = 110$ GeV) indicate the GC best fit pMSSM models from ref.~\cite{Caron:2015wda}, which coincide with the best global fit models obtained by~\cite{2015arXiv150707008B}.  The dotted red line indicates the $1\sigma$ contour of the most likely pMSSM10 models from ref.~\cite{Bagnaschi:2015eha} (only for wino NLSP). Figures (a) and (b) are made using a minimal jet $p_T$ of 50 GeV. Figures (c) and (d) using the ATLAS trigger lepton $p_T$ requirements as shown in table~\ref{tab:searches}.}
\end{figure}

\addcontentsline{toc}{section}{Appendix: Detailed discussion of the LHC14 reach}
In this appendix we show what happens if different assumptions on the cuts, the systematic error and the integrated luminosity are made. In the standard analysis, a jet veto with $p_T(j) < 30$ GeV is used. Using this cut, the LHC operating at 14 TeV and with 300~fb$^{-1}$ of integrated luminosity can probe LSP masses up to 140~GeV for mass gaps between $\sim 9-50$~GeV if the NLSPs are wino-like and LSP masses up to 95~GeV for $\Delta m \gtrsim 20$ GeV if the NLSPs are higgsino-like. If a jet veto with $p_T(j) < 50$~GeV is used (figures~\ref{fig:winoatl50} and~\ref{fig:hinoatl50}), the LHC can probe LSP masses up to 135~GeV~(85~GeV) for $\Delta m \gtrsim$ 10 GeV (25 GeV) for the wino (higgsino) NLSP region. \\

In the standard analysis, the requirements on the transverse momentum of the leptons are $5 $ ${\rm GeV} < p_T(\mu) < 50$~GeV for muons and $10$ $ {\rm GeV} < p_T(e) < 50$~GeV for electrons. In figure~\ref{fig:winoatl30} (\ref{fig:hinoatl30}) we show the reached significance using the current ATLAS lepton trigger $p_T$ requirements (as indicated in table~\ref{tab:searches}) for the wino (higgsino) NLSP region. The significance is somewhat reduced compared to the significance using lower requirements on the lepton transverse momenta. Using the ATLAS trigger lepton $p_T$ requirements, we can exclude wino pMSSM scenarios with LSP masses up to 135 GeV and mass gaps $\gtrsim 10$ GeV. Therefore, if it is possible to lower the lepton transverse momentum trigger requirements, this would be worth to pursue. \\

\begin{figure}[h]
 \subfigure[]{\label{fig:winopt5001}\includegraphics[width=0.485\textwidth]{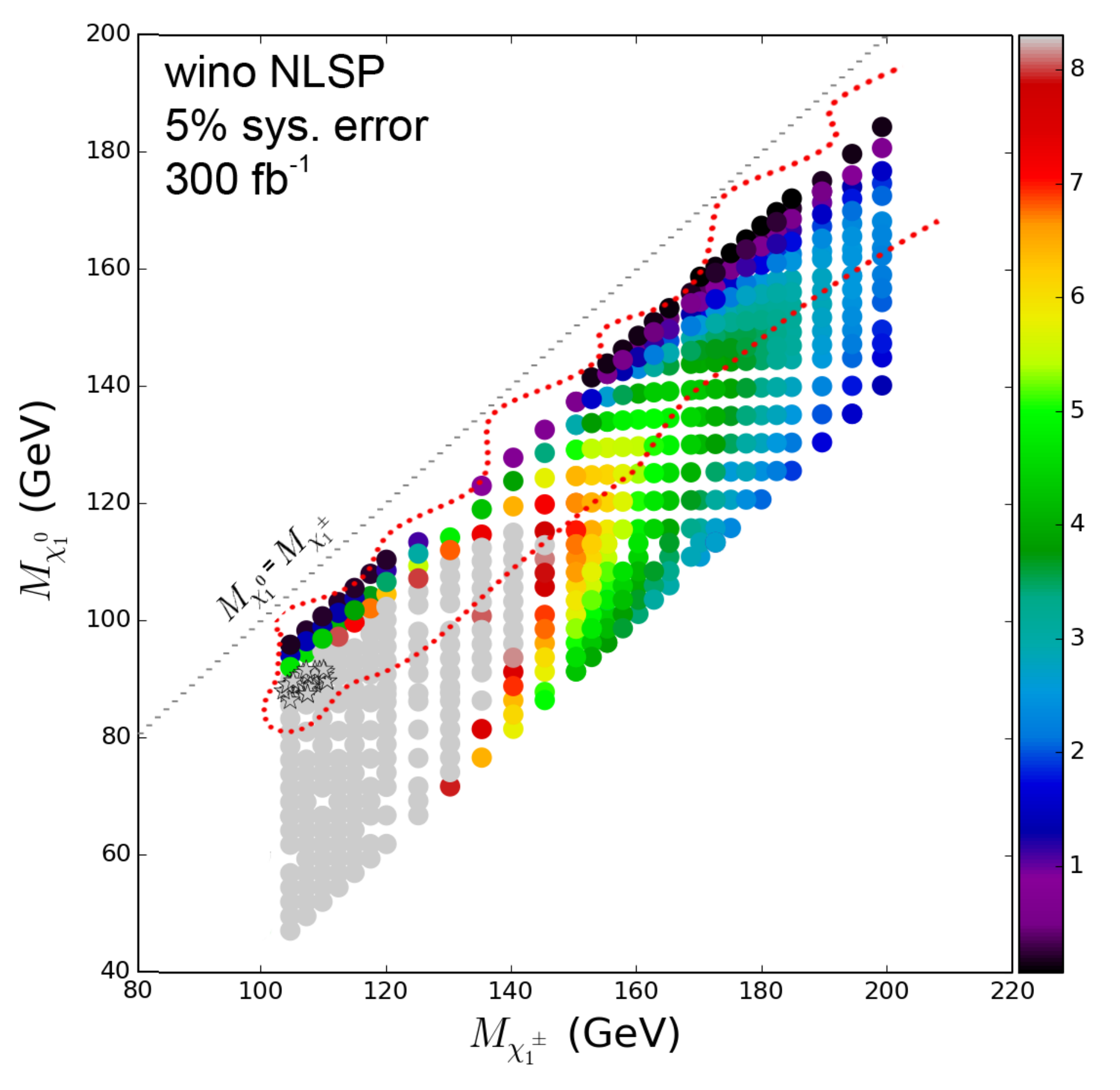}}
 \subfigure[]{\label{fig:hinopt5001}\includegraphics[width=0.485\textwidth]{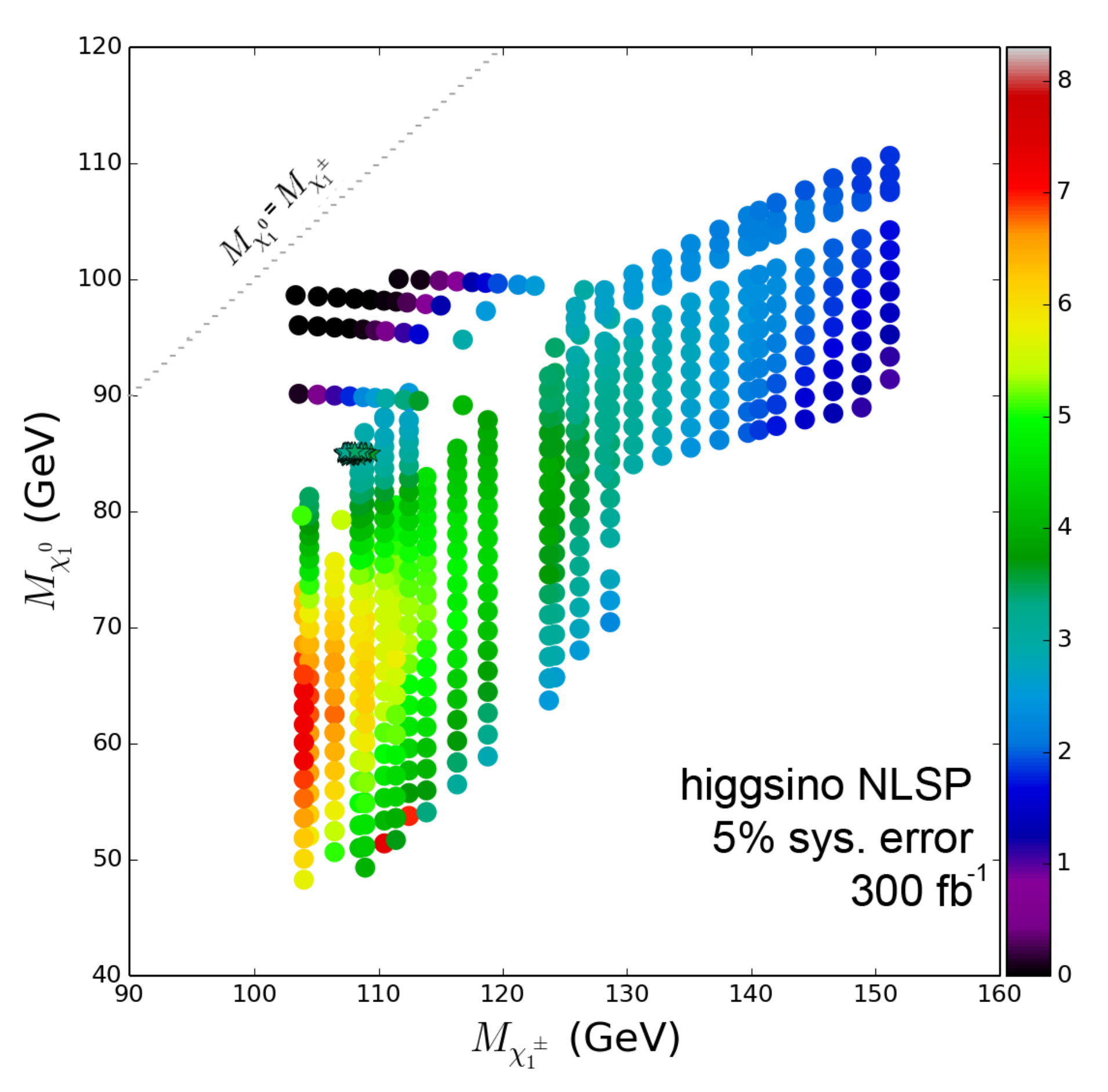}}

 \subfigure[]{\label{fig:wino30}\includegraphics[width=0.485\textwidth]{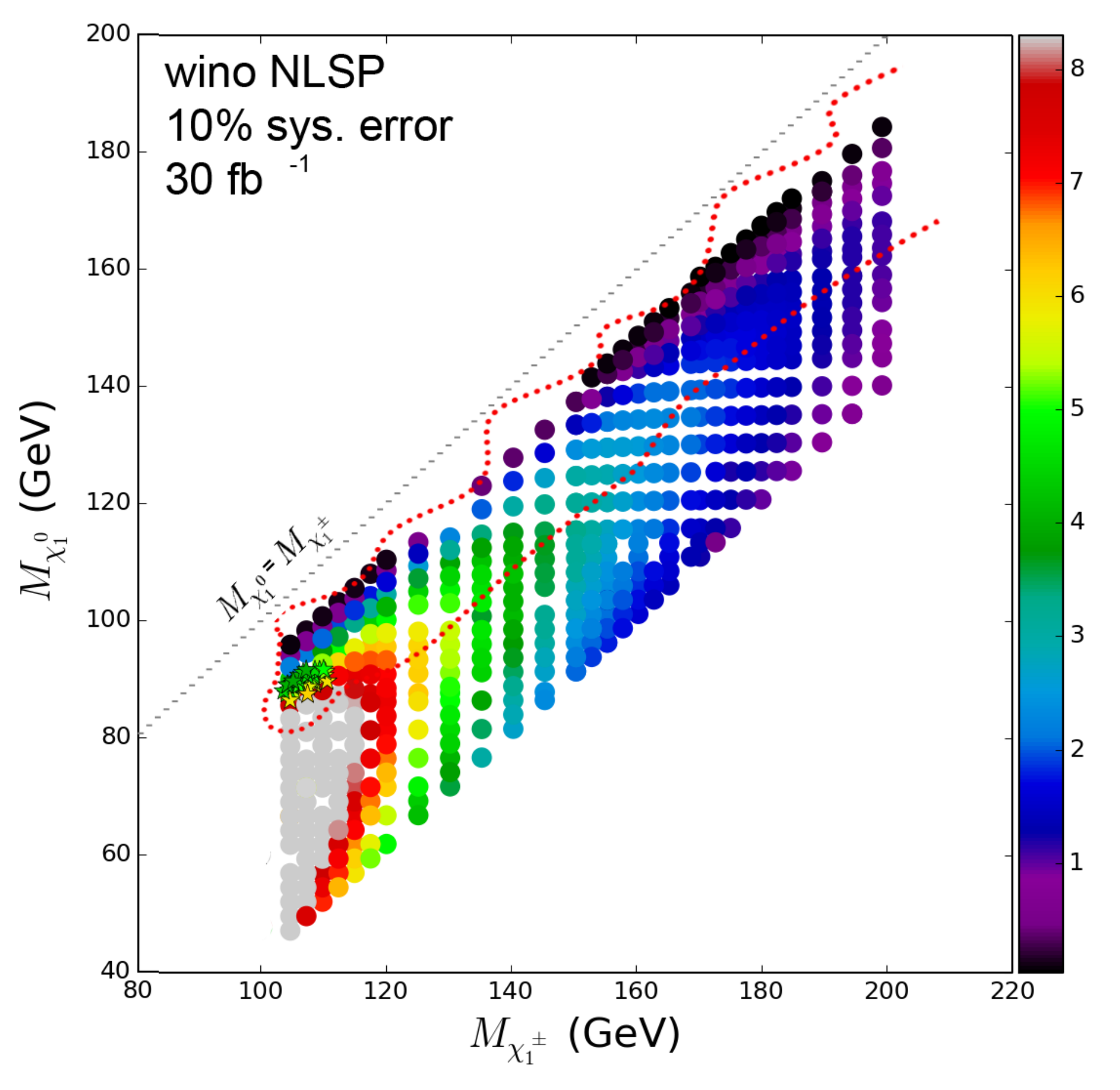}}
\subfigure[]{\label{fig:hino30}\includegraphics[width=0.485\textwidth]{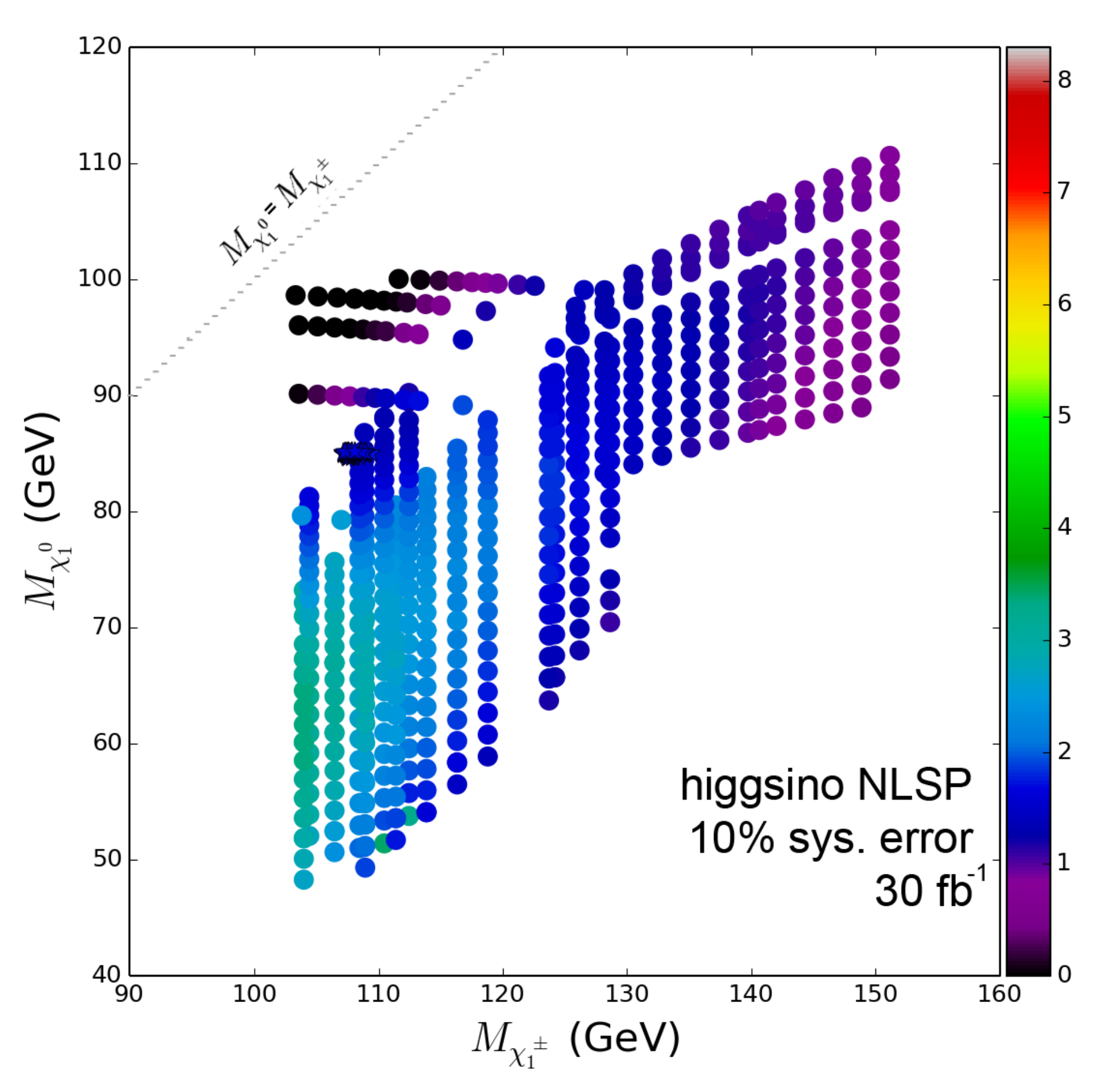}}
\caption{Significance (indicated by the color scale) for wino NLSP models (\emph{left}) and higgsino NLSP models (\emph{right}). Stars (located around $m_{\tilde{\chi}^{\pm}_1} = 110$ GeV) indicate the GC best fit pMSSM models from ref.~\cite{Caron:2015wda}, which coincide with the best global fit models obtained by~\cite{2015arXiv150707008B}.  The dotted red line indicates the $1\sigma$ contour of the most likely pMSSM10 models from ref.~\cite{Bagnaschi:2015eha} (only for wino NLSP). Figures (a) and (b) are made assuming a systematic background uncertainty of 5$\%$.  Figures (c) and (d) are made assuming an integrated luminosity of 30 fb$^{-1}$. }
\end{figure}

\begin{figure}[h]
 \subfigure[]{\label{fig:winoatl5}\includegraphics[width=\textwidth]{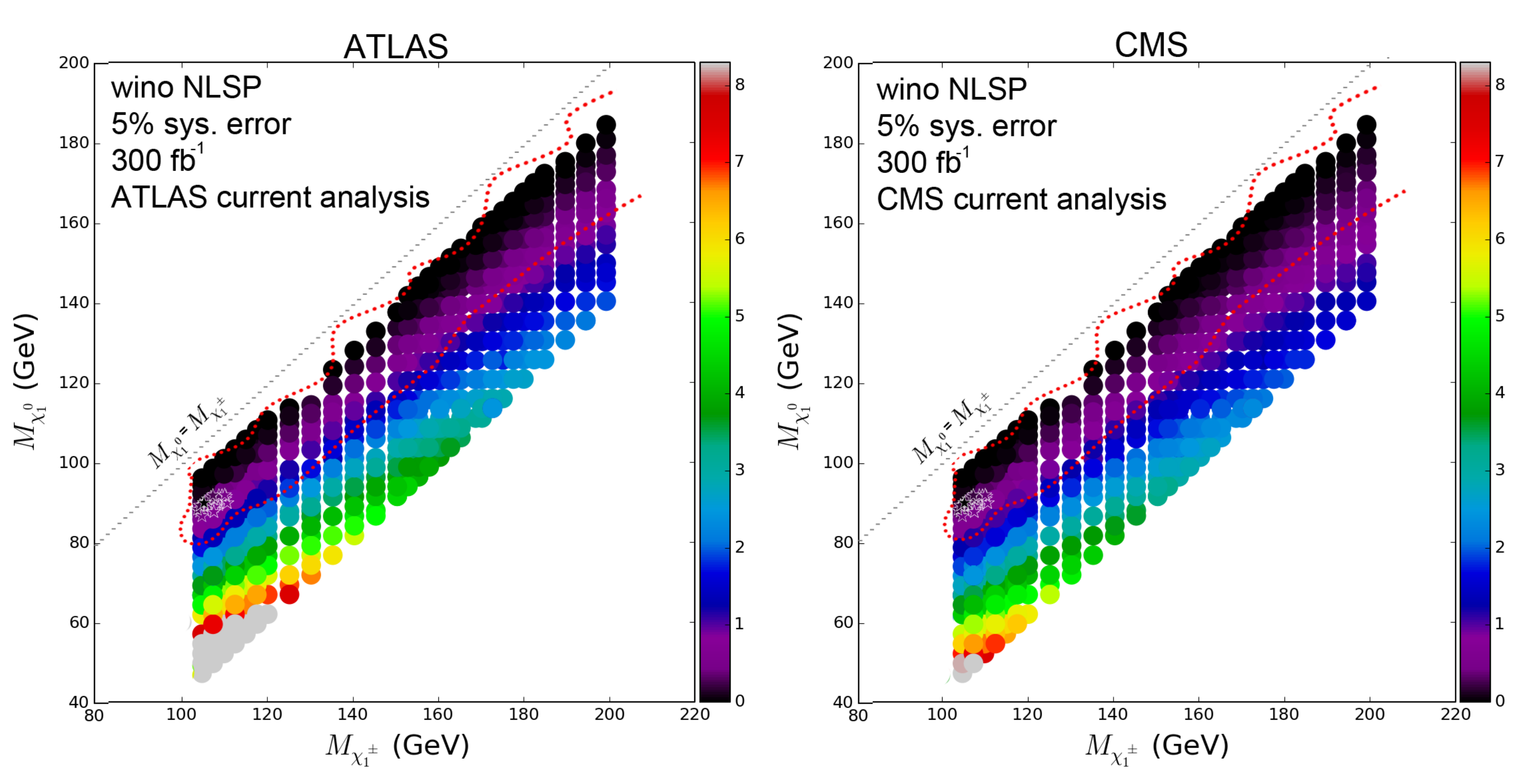}}
 \subfigure[]{\label{fig:winoatl3000}\includegraphics[width=\textwidth]{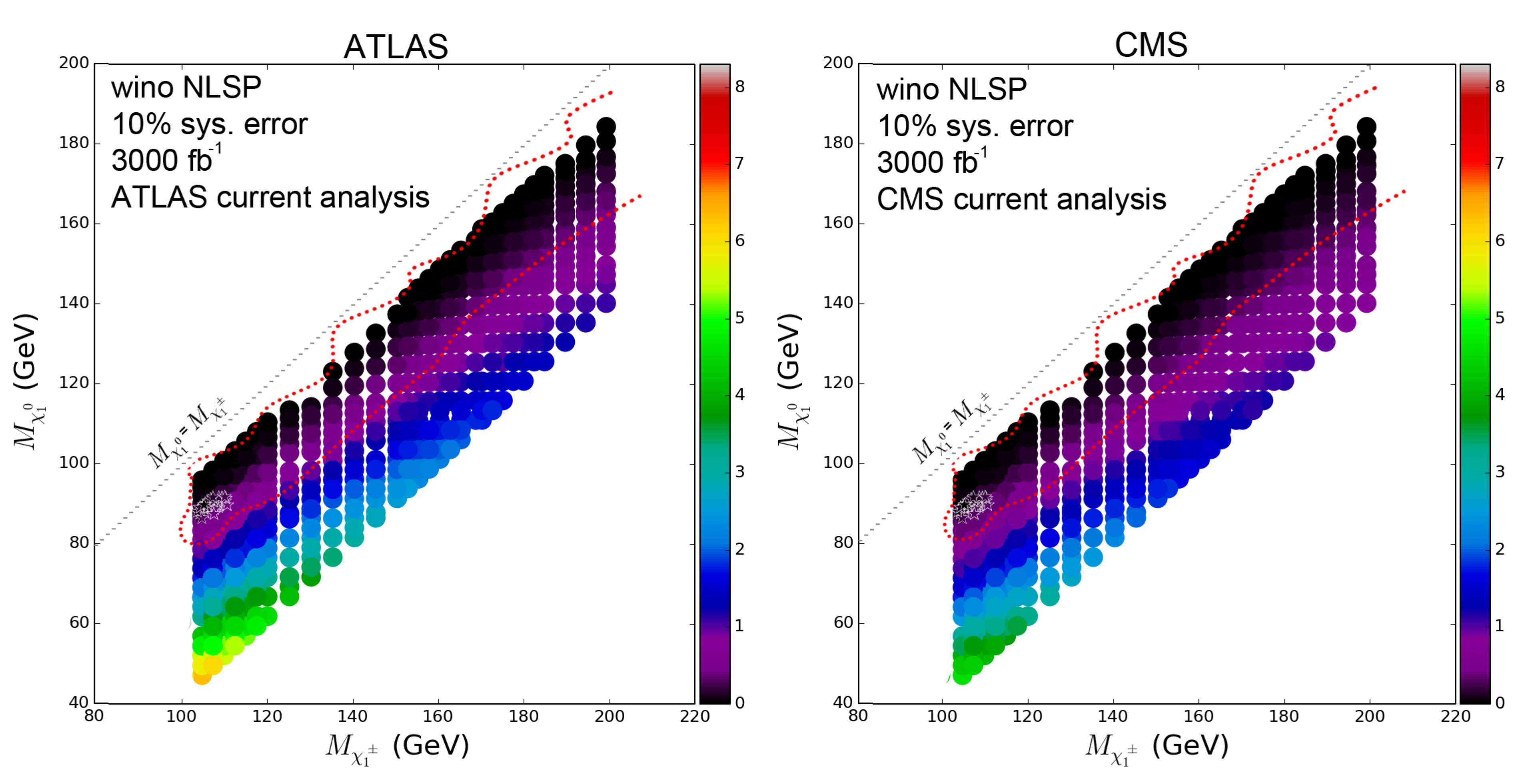}}
\caption{Standard ATLAS and CMS tri-lepton search significances (indicated by the color scale) for wino NLSP pMSSM models assuming (a) 300 fb$^{-1}$ and 5$\%$ systematic error and (b) 3000 fb$^{-1}$ and 10$\%$ systematic error. The dashed gray line indicates the limit $m_{\tilde{\chi}^0_1} = m_{\tilde{\chi}^{\pm}_1}$. Stars (located around $m_{\tilde{\chi}^{\pm}_1} =~110$~GeV) indicate the GC best fit pMSSM models from ref.~\cite{Caron:2015wda}, which coincide with the best global fit models obtained by~\cite{2015arXiv150707008B}.  The dotted red line indicates the $1\sigma$ contour of the most likely pMSSM10 models from ref.~\cite{Bagnaschi:2015eha}. }
\vspace{-1cm}
\end{figure}

\begin{figure}[h]
 \subfigure[]{\label{fig:hinoatl5}\includegraphics[width=\textwidth]{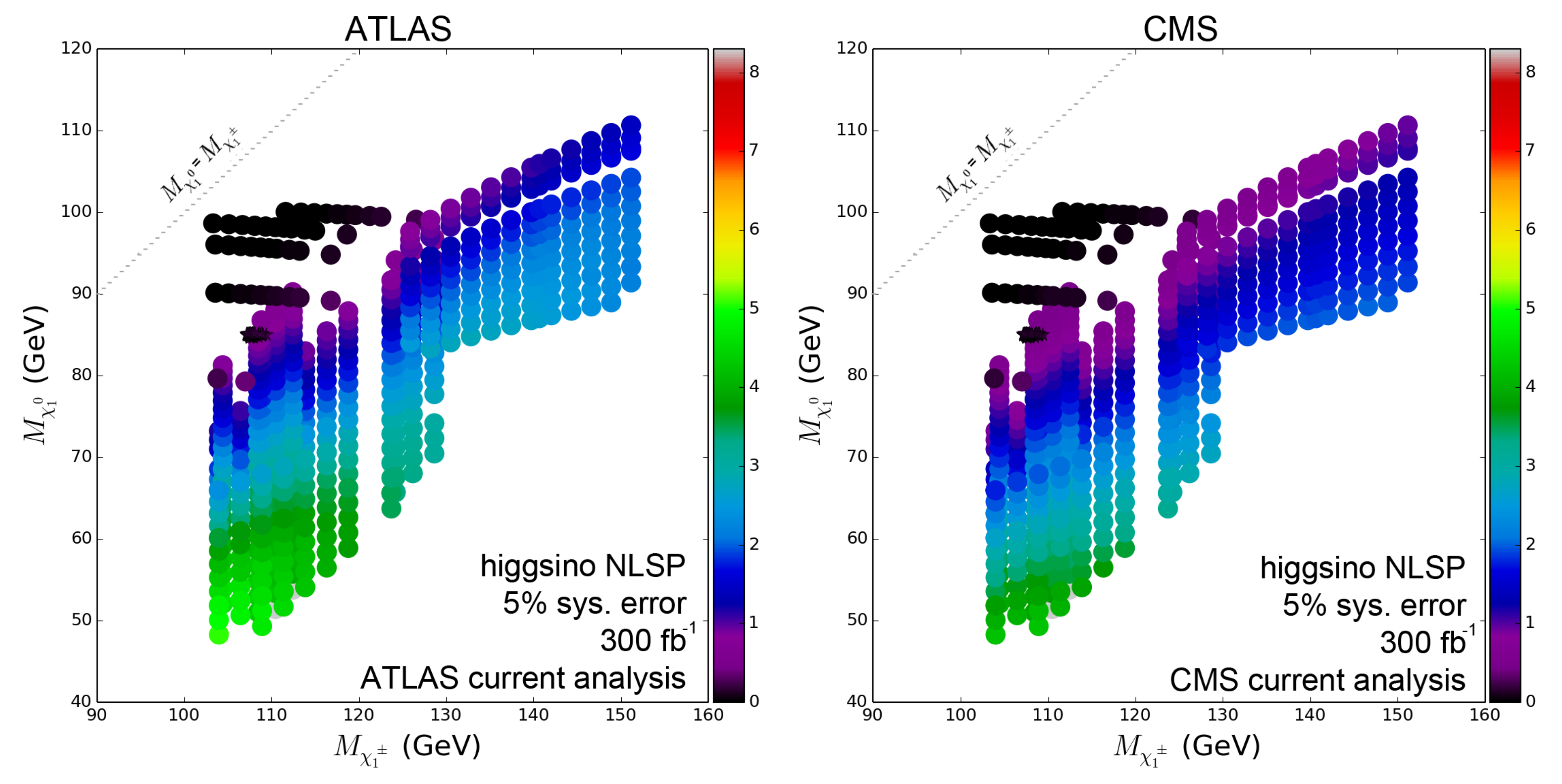}}
 \subfigure[]{\label{fig:hinoatl3000}\includegraphics[width=\textwidth]{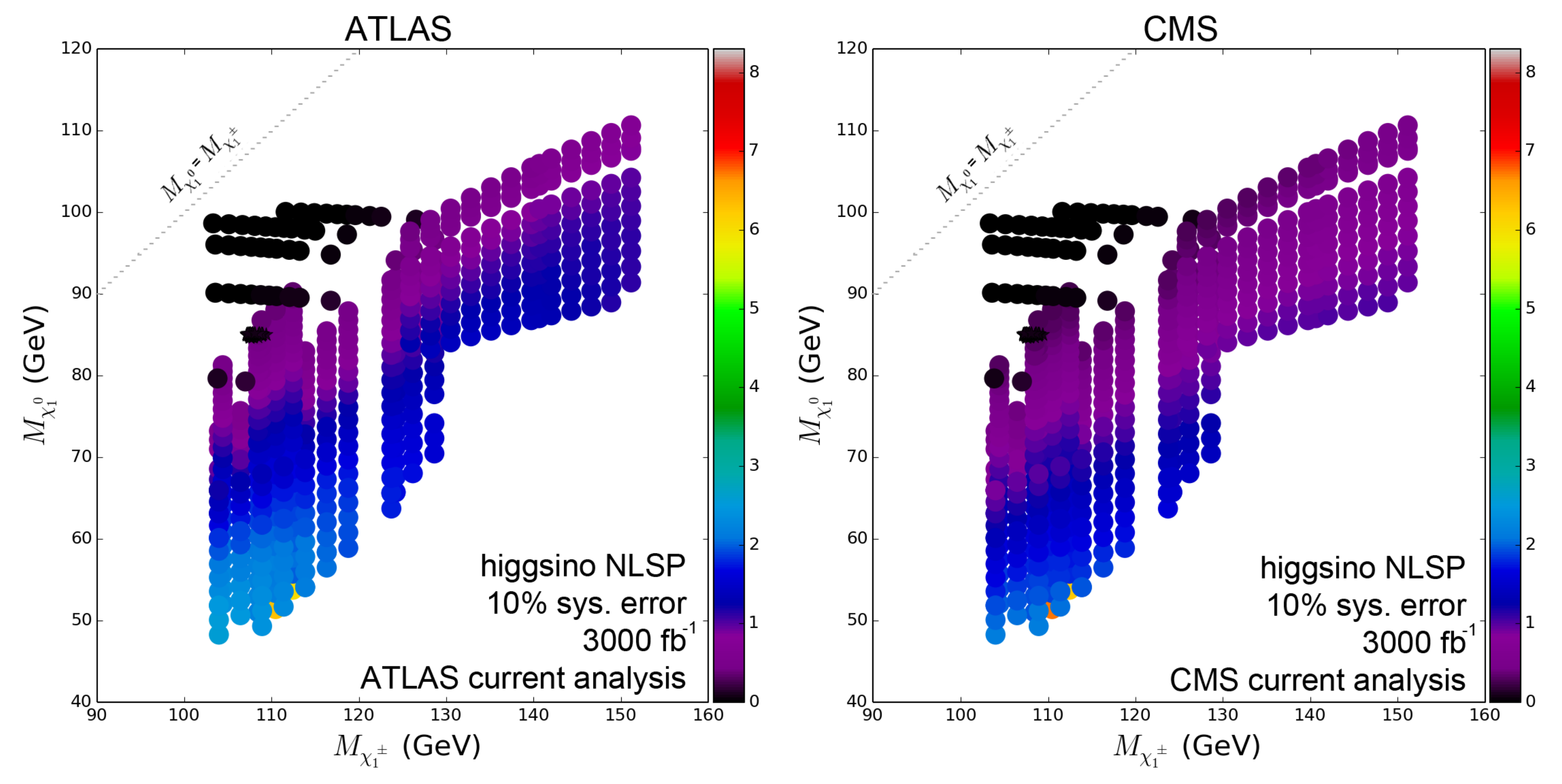}}
\caption{Standard ATLAS and CMS tri-lepton search sensitivities (indicated by the color scale) for higgsino NLSP pMSSM models assuming (a) 300 fb$^{-1}$ and 5$\%$ systematic error and (b) 3000 fb$^{-1}$ and 10$\%$ systematic error. The dashed gray line indicates the limit $m_{\tilde{\chi}^0_1} = m_{\tilde{\chi}^{\pm}_1}$. Stars (located around $m_{\tilde{\chi}^{\pm}_1} =~110$~GeV) indicate the GC best fit pMSSM models from ref.~\cite{Caron:2015wda}, which coincide with the best global fit models obtained by~\cite{2015arXiv150707008B}.}

\end{figure}

If the systematic error could be reduced to 5$\%$ (figure~\ref{fig:winopt5001} and~\ref{fig:hinopt5001}), the sensitivity would be greatly enhanced. In that case, exclusion of LSP masses $>170$ ($110$) GeV with mass gaps $\gtrsim6$ (15) GeV can be realized for the wino (higgsino) NLSP pMSSM models. In contrast, the ATLAS and CMS experiments with their current tri-lepton search strategies are in that case still not able to probe the favored pMSSM regions (figure~\ref{fig:winoatl5} and~\ref{fig:hinoatl5}). \\

Even using an integrated luminosity of 30 fb$^{-1}$, the proposed analysis can exclude the wino (higgsino) NLSP scenarios for LSP masses up to 135 (85) GeV and $\Delta m \gtrsim 10$ (25) GeV (figures~\ref{fig:wino30} and~\ref{fig:hino30}). Using the proposed cuts and assuming a systematic background uncertainty of 10$\%$, we can be sensitive to the 100 GeV bino-like DM particle with an integrated luminosity of 30 fb$^{-1}$, while the ATLAS and CMS experiments with their current tri-lepton searches are not even sensitive with an integrated luminosity of 3000 fb$^{-1}$ (figures~\ref{fig:winoatl3000} and~\ref{fig:hinoatl3000}). \\

We therefore conclude that an updated search strategy is needed. Even in the most optimistic cases (using 3000 fb$^{-1}$ of data or reducing the systematic error on the standard model background to 5$\%$) the current tri-lepton search strategies of ATLAS and CMS are not sensitive to the discussed favored regions of the pMSSM parameter space.


\clearpage
\bibliographystyle{jhep} 
\bibliography{bibreport}
\end{document}